\documentclass[aps, pra, a4paper, showpacs, twocolumn, english, nofootinbib, 10pt, longbibliography]{revtex4-1}
\usepackage{bbm, amsthm, bm, textcomp, nicefrac, geometry, ragged2e}
\usepackage[dvipsnames]{xcolor}
\usepackage{float}
\usepackage[bbgreekl]{mathbbol}
\usepackage{graphicx, epstopdf, color, verbatim, enumitem}
\usepackage{pbox,array}                                                 
\usepackage{mathrsfs}
\usepackage{upgreek}
\usepackage{amssymb}
\usepackage{lipsum}
\usepackage{textgreek}
\usepackage{mathtools}
\usepackage{stackrel}
\usepackage[thinlines]{easytable}
\usepackage{amsmath}
\usepackage[utf8]{inputenc}
\makeatletter
\usepackage{soul}
\usepackage[caption=false]{subfig}
\usepackage[colorlinks]{hyperref}
\hypersetup{
    colorlinks = true,
    linkcolor = Purple,
    citecolor = Purple,
    filecolor = Purple,
    urlcolor = Purple}
\usepackage[T1]{fontenc}
\usepackage[caption=false]{subfig}
\usepackage{babel}
\usepackage[normalem]{ulem}
\usepackage[linesnumbered,ruled]{algorithm2e}

\addto\captionsenglish{}

\newcommand\id{\mathbbm{1}}
\newcommand\ti{\text{i}}
\newcommand{\ket}[1]{\left| #1 \right\rangle}
\newcommand{\bra}[1]{\left\langle #1 \right|}
\newcommand{\braket}[2]{\left\langle #1 | #2 \right\rangle}
\newcommand{\ketbra}[2]{\left| #1\right\rangle\!\left\langle#2\right|}
\newcommand{\proj}[1]{\left| #1\right\rangle \! \left\langle #1 \right|}
\newcommand{\bo}[1]{\boldsymbol{#1}}

\definecolor{brickred}{rgb}{0.8, 0.0, 0.0}

\begin{document}

\title{Quantum computation with logical gates between hot systems}

\author{Ferran Riera-S\`{a}bat$^1$, Pavel Sekatski$^2$, and Wolfgang D\"{u}r$^1$}

\affiliation{$^1$Universit\"{a}t Innsbruck, Institut f\"{u}r Theoretische Physik, Technikerstra{\ss}e 21a, Innsbruck 6020, Austria \\
$^2$University of Geneva, Department of Applied Physics, Geneva 1211, Switzerland}

\date{\today}

\begin{abstract}
We consider quantum computer architectures where interactions are mediated between hot qubits that are not in their mechanical ground state. Such situations occur, e.g., when not cooling ideally, or when moving ions or atoms around. We introduce quantum gates between logically encoded systems that consist of multiple physical ones and show how the encoding can be used to make these gates resilient against such imperfections. We demonstrate that, in this way, one can improve gate fidelities by enlarging the logical system, and counteract the effect of unknown positions or position fluctuations of involved particles. We consider both a classical treatment of positions in terms of probability distributions, and a quantum treatment using mechanical eigenmodes. We analyze different settings including a cool logical system mediating interactions between two hot systems, as well as two logical systems consisting of hot physical systems whose positions fluctuate collectively or individually. In all cases, assuming ideal local control to logical systems, we demonstrate a significant improvement in gate fidelities, which provides a platform-independent tool to mitigate thermal noise in the context of trapped-particle-based architectures.
\end{abstract}

\maketitle

\section{Introduction}

Quantum computers offer the promise to enhance the efficiency to solve various problems, or even enlarge the class of accessible ones. Several architectures to design large-scale quantum computers exist \cite{hot_o2007optical, hot_monroe2013scaling, hot_brecht2016multilayer}. While some approaches are intrinsically scalable, others rely on the combination of small modules that can be connected \cite{hot_moudlarMonroe2014, hot_Lekitsch_2017, hot_akhtar2023high}. Promising platforms include trapped ions in segmented traps, where ions are shuttled around to an interaction zone \cite{hot_kielpinski2002architecture, hot_Home_2009_shuttlingions, hot_pino2021demonstration}. Similarly, arrays of trapped Rydberg atoms have been realized, where atoms can be moved using optical tweezers and interact via induced dipole-dipole interactions \cite{hot_demille2002quantum, hot_yelin2006schemes, hot_Browaeys_2016, hot_Henriet2020quantumcomputing, hot_bluvstein2022quantum, hot_graham2022multi}. But also other types of segmented traps are conceivable, where, e.g., 1D or 2D arrays of ions are manipulated by laser pulses \cite{hot_ruster2014experimental, hot_scholl2021quantum}, and where interaction between ions in different modules takes place via some distance-dependent coupling \cite{hot_Porras2004, hot_Joshi_2020, hot_Pagano_2020, hot_wan2020ion, hot_monroe2021programmable}. What most approaches have in common is the necessity to cool particles to their mechanical ground state, to enable their manipulation and gates between them with high fidelity \cite{hot_rowe2002transport, hot_barrett2003sympathetic, hot_blakestad2009high, hot_bowler2012coherent, hot_ruster2014experimental, hot_brown2016co}. For different platforms, explicit approaches have been developed to perform gates between hot systems, and avoid or at least reduce the demand for cooling \cite{hot_Leandro_2007, hot_cirac_hotqubits_1998, hot_sorensen_1999}.

Here we present a novel, generic, and platform-independent approach to deal with the influence of thermal fluctuations and position noise in particle-trap-based architectures. We consider a setting where qubit systems interact via some distance-dependent commuting coupling \cite{hot_review-long-range-interactions}, which can be, e.g., induced by laser pulses \cite{hot_Porras2004, hot_richerme_2014non, hot_zhang2017observation, hot_Joshi_2020, hot_Pagano_2020} or via dipole-dipole interactions \cite{hot_demille2002quantum, hot_yelin2006schemes, hot_Browaeys_2016, hot_bluvstein2022quantum}. Obviously, such an approach is susceptible to position noise and thermal fluctuations. Rather than attempting to cool the system, we use a logical encoding of quantum information to tailor effective interactions and make them insensitive to particle positions. The basic idea is to encode quantum information in multiple physical systems, that form one logical qubit \cite{hot_Devitt_2013, hot_RevModPhys.87.307, hot_Dur_2007, hot_Bultrini2023battleofcleandirty, hot_koukoulekidis2023framework}. The physical systems interact, which generates an effective interaction between a logical system and other physical ones, or between two logical systems. By properly choosing the encoding, we show that thermal noise can be significantly suppressed. Increasing the size of encoding, i.e., the number of involved physical systems, leads to larger gate fidelities that approach unity.

We demonstrate the applicability of our approach in multiple setups, which include classical treatment of particle positions in terms of trajectories or probability distributions, as well as a full quantum treatment using mechanical eigenmodes for ions trapped in a 1D Paul trap:
\begin{enumerate}
    \item[a)] We show that one can mediate interactions and gates between two hot physical systems that suffer from thermal fluctuations utilizing a cool logical system.
    \item[b)] We consider collective fluctuations, where relative positions of the individual systems in the same logical system are fixed, while the position of each logical system is given by a probability distribution, e.g., a Gaussian. Logical systems are either 1D chains or 2D arrays of trapped particles.
    \item[c)] We treat two logical systems where each particle is independently affected by classical thermal noise.
    \item[d)] We consider a fully quantized model of 1D Paul traps and their mechanical eigenstates, demonstrating that interactions between two logical ion strings can be made insensitive to their temperature.
\end{enumerate}

\begin{figure}
    \centering
    \subfloat[]{\includegraphics[width=0.47\columnwidth]{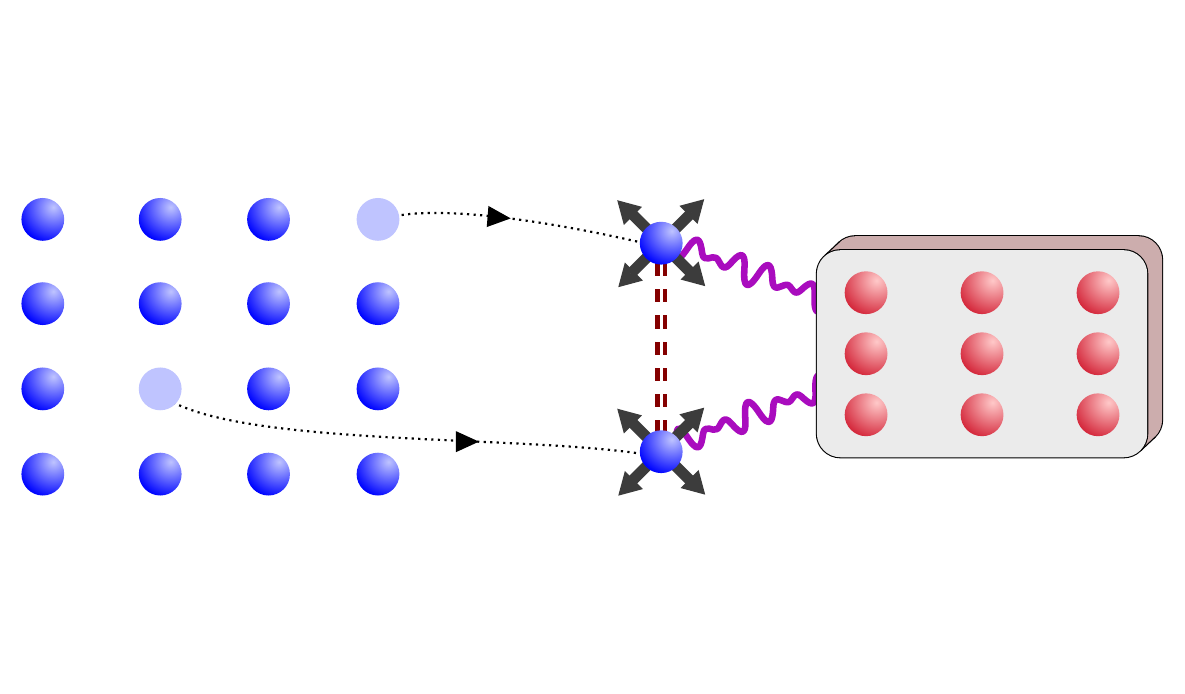} \label{fig.model1}} \hspace{0.2cm}
    \subfloat[]{\includegraphics[width=0.47\columnwidth]{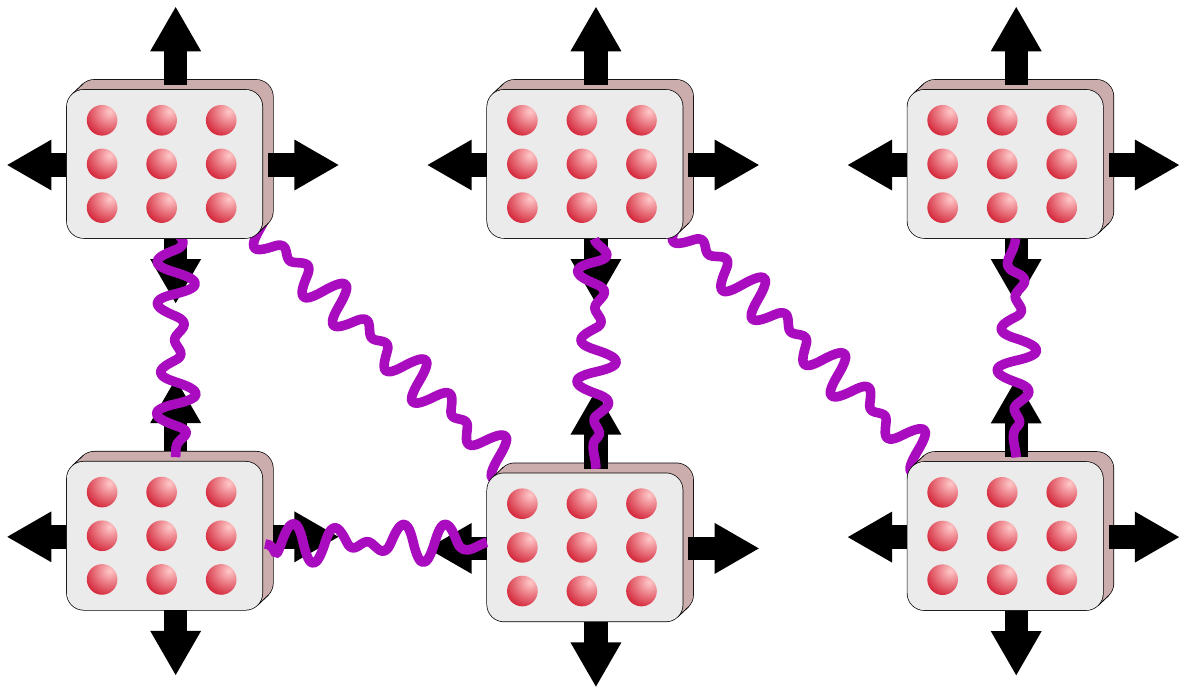}\label{fig.model2}} \\
    \subfloat[]{\includegraphics[width=0.47\columnwidth]{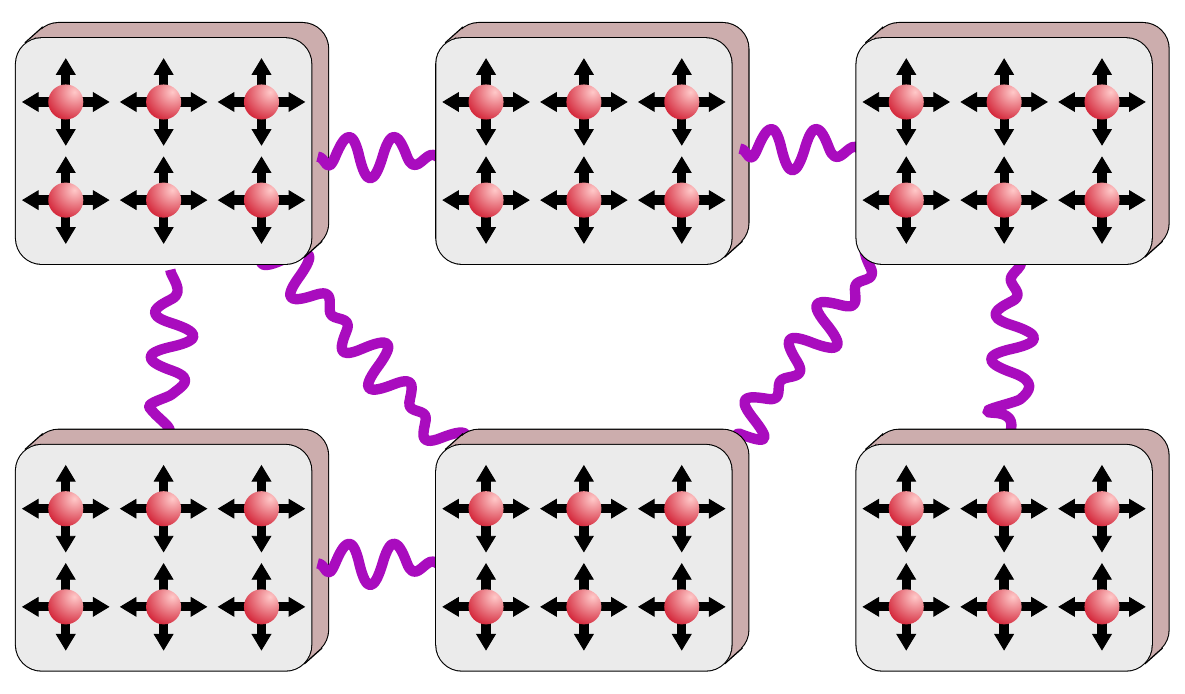}\label{fig.model3}} \hspace{0.2cm}
    \subfloat[]{\includegraphics[width=0.47\columnwidth]{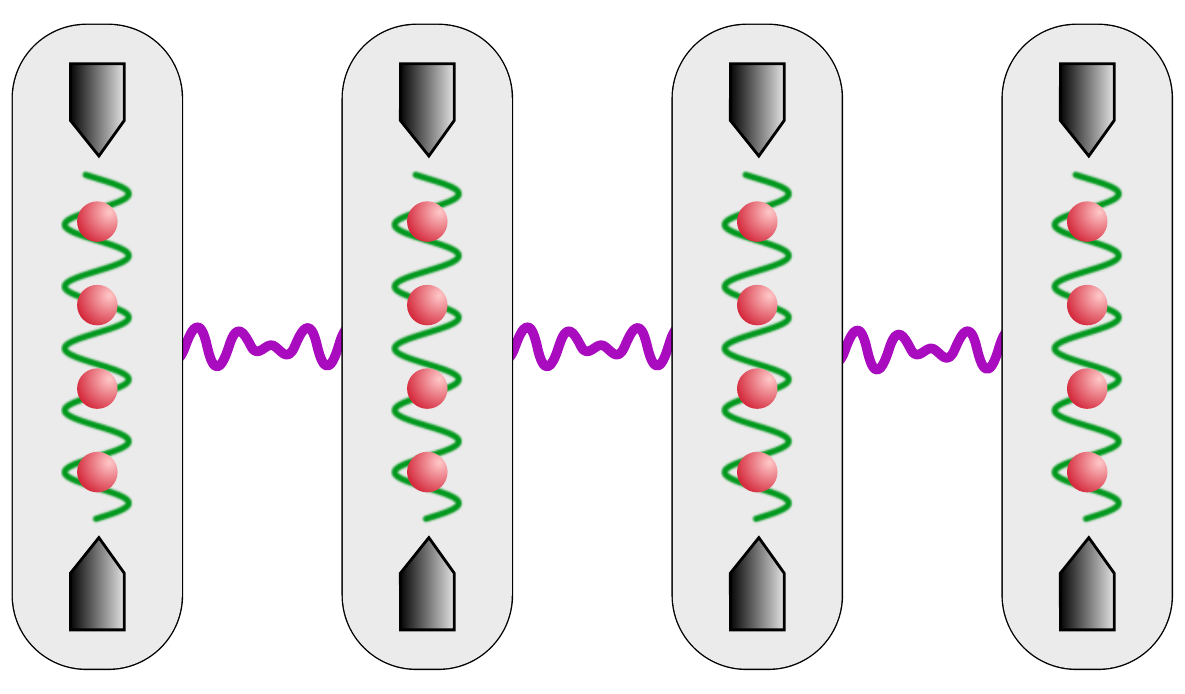}\label{fig.model4}}
    \caption{\label{fig:models} (a) The data qubits are moved close to the mediator module, which mediates an effective interaction between them. The data qubits are affected by position noise. (b,c,d) Each module encodes a logical system and they couple with each other via the inherent interaction between the physical systems. (b) The positions of the physical qubit systems fluctuate collectively. (c) The position of each physical qubit system fluctuates independently. (d) Within each module, the position of each physical qubit system is quantized and correlated with the others.}
\end{figure}

The considered setups are illustrated in Fig.~\ref{fig:models}, even though when analyzing different cases we will restrict the position noise to dominant directions or degrees of freedom. In all cases, we find a significant improvement in achievable gate fidelities. The situation of a) is relevant when shuffling ions or atoms around, which are heated up due to this process \cite{hot_brown2016co, hot_sutherland2021motional}. They can be moved to some interaction zone, where a properly cooled system consisting of several qubits is used to mediate an interaction between the two hot particles. Similarly, this can account for the heating of ions or atoms due to measurements, or the application of multiple gates. The setup of b) consists of two independent 1D or 2D traps, where particles interact due to some induced, distant-dependent coupling \cite{hot_Porras2004, hot_Joshi_2020, hot_Pagano_2020, hot_wan2020ion, hot_monroe2021programmable}. The setting considered in c) is, e.g., concerned with independent traps for each of the constituents, and hence independent thermal fluctuations. It may also relate to situations where all systems are moved and hence heated up. Finally, the fully quantized version we treat in d) is concerned with two independent ion strings in Paul traps, where again some distant-dependent coupling between the ions is induced to couple the two traps. We consider the Boltzmann distribution of the mechanical energy eigenstates and compute resulting gate fidelities when increasing the number of ions in each trap.

Due to the kind of encoding we use, our approach can at the same time be used as an error correction code against bit-flip errors. This however requires additional control and overhead but may be useful in a quantum computation setup that goes beyond the usage of bare qubits. In addition, we can modify our scheme such that it also protects against fluctuations of a constant background field, one of the main sources of decoherence in certain ion-trap setups. Similarly, fluctuations against background fields and noise sources with specific spatial dependence or fixed origin can be achieved. 

The paper is organized as follows. In Sec. \ref{sec.Quan.computers.models} we describe different quantum computation setups, as well as the general setting and approach we consider. In Sec. \ref{sec.From.the.physical.to.a.logical.layer} we introduce how we model thermal noise and how we can utilize logical encodings to mitigate the noise effects. In Secs. \ref{sec.Cold.mediating.system}-\ref{sec.position.quantitzation} we discuss the different settings (a)-(d), and provide explicit examples that demonstrate the performance of our approach. In Sec. \ref{sec: background noise} we briefly discuss how to achieve protection against a noisy background field, without increasing the size of the logical system. We summarize and conclude in Sec. \ref{sec.conclusions}.

\section{Modular quantum computer models}
\label{sec.Quan.computers.models}

The primary strategy for developing a scalable quantum computer is based on modularity \cite{hot_moudlarMonroe2014, hot_Lekitsch_2017, hot_akhtar2023high}. This approach involves constructing the entire system by interconnecting smaller, independent quantum processors that can execute quantum operations and store quantum information. One approach to interconnecting modules relies on distant-dependent interactions among the constituent physical systems. However, by definition, such interactions are sensitive to position fluctuations. Hence, achieving a high-fidelity interaction requires cooling down the systems to their mechanical ground state, which is not always feasible. In this paper, we introduce an alternative (or complementary) solution, demonstrating how the adverse effects of mechanical thermal noise can be mitigated by leveraging the interaction between logical systems encoded within multiple physical ones.

In a multiqubit system, one can implement a logical qubit by restricting the state of the system into a two-dimensional subspace. In particular, we consider the repetition encoding where the logical computational basis is of the form $\ket{\bar{0}} = \ket{\bo{a}}$ and $\ket{\bar{1}} = \ket{-\bo{a}}$, where $\ket{\bo{a}}$ is a state of the computational basis such that $Z_i \ket{\bo{a}}  = a_i \ket{\bo{a}}$. We refer to $\bo{a}$ as the logical vector and as we show in the next section, interactions between logical qubits can be made approximately position-independent by a proper choice of $\bo{a}$. In principle, $a_i \in \{ -1, 1\}$, however by performing fast flips of the individual qubits during the evolution we can effectively obtain an arbitrary logical vector with $a_i \in [-1, 1]$ (see Sec. 3.6 in Ref. \cite{hot_riera2023simulator}). Next, we describe several modules-connected-based architectures subjected to different kinds of thermal noise and how logical encoding can be used to mitigate the effects of position fluctuations.

\subsection{Interactions mediator system}
\label{sec.Interactions.mediator.system}

The first model we consider resembles the von Neumann architecture \cite{hot_von_Neumann_architecture} and is based on an auxiliary system (or module) that is used to mediate interactions between hot physical systems \cite{hot_burd2021quantum}. We consider two kinds of modules. The data module consists of multiple trapped physical systems, e.g., ions or neutral atoms, where each encodes a qubit state. We assume the state of the qubits can be individually manipulated but multiqubit gates cannot be directly implemented. For that, the physical systems are individually moved into an interaction zone \cite{hot_kielpinski2002architecture, hot_Home_2009_shuttlingions, hot_pino2021demonstration} where they interact with an auxiliary system, which is used to mediate interactions between the data qubits \cite{hot_riera2023remotely}; see Fig.~\ref{fig.model1}. For instance, the multiqubit gate $e^{\ti \alpha Z^{\otimes n}}$ can be mediated on the data qubits by applying a control-$Z$ gate between each data qubit and the auxiliary system, and then measuring the auxiliary system on the appropriate basis, i.e.,
\begin{equation}
\begin{aligned}
\label{eq:mediating}
    (Z^{\otimes n})^k_{1 \dots n} \, P^{(k)}_0 \, e^{\ti \alpha X_0} \, \prod_{j=1}^n \text{CZ}_{0j} \ket{+}_0 \ket{\psi}_{1\dots n} & \\
    \mapsto \, e^{\ti \alpha Z^{\otimes n}}\ket{\psi}_{1\dots n} & ,
\end{aligned}
\end{equation}
where $P^{(k)} = \proj{k}$, and $\ket{\psi}$ is an arbitrary state. In this model, the mediator module also consists of multiple trapped physical systems but it is assumed to be well cooled and fully controllable. By encoding a logical qubit in the mediator module, we establish interactions that mitigate the effects of thermal noise acting on the components of the data module.

In Sec.~\ref{sec.Cold.mediating.system} we analyze the fidelity of a mediated interaction by computing the fidelity of a control gate between a noiseless logical qubit system and a hot physical qubit system.

\subsection{Independent classical modules}
\label{sec.Independent.classical.modules}

The second scheme is based on independent modules consisting of multiple physical qubit systems which can be fully controlled. The modules are coupled with each other by a distance-dependent interaction between their physical components. We assume the modules are affected by collective position noise, meaning that the physical systems are well trapped within the modules but the position of the latter is subject to position fluctuations; see Fig.~\ref{fig.model2}. In this case, we implement a logical qubit within each module and make use of their interactions. Similar to the previous scheme, logical systems allow us to establish high-fidelity interactions. In Sec.~\ref{sec.Collective.position.noise} we analyze how the fidelity of the interaction between two modules can be enhanced by enlarging the size of the modules.

\subsection{Independent physical qubit systems}
\label{sec.Independent.physical.qubits}

The third scheme is similar to the one considered in the previous section, but in this case, thermal noise affects all physical systems independently, see Fig.~\ref{fig.model3}. This scenario is of particular interest as the total amount of noise affecting a module increases with its size. However, in Secs.~\ref{sec.Independent.noise} and \ref{sec.General.case} we show, both in a classical and quantized approach, that grouping the qubits in logical systems still allows one to arbitrarily enhance the interaction by increasing the system size. In this case, alternative ways of coupling the qubits within the modules should be used as the independent nature of noise would affect the manipulation of the logical systems.

\subsection{Independent quantum modules}
\label{sec.Independent.quantum.modules}

The last scheme we consider is a quantized version of independent modules. For example, each module is a 1D quantum trap where at a certain temperature the collective motion of the physical systems is described by the thermal state over their mechanical eigenmodes; see Fig.~\ref{fig.model4}. These fluctuations lead to a noisy interaction between different traps. In contrast to the classical model introduced in Sec.~\ref{sec.Independent.classical.modules}, the noise impact of the whole system grows with the number of trapped particles. However, the temperature does not affect the physical systems independently but instead excites the collective oscillation modes of the trap, which we take as an advantage. In Sec.~\ref{sec.traps}, we show that by implementing logical systems with trapped physical systems one can obtain significant gate fidelity enhancements.
 
\section{From the physical to a logical layer}
\label{sec.From.the.physical.to.a.logical.layer}

In this section, we introduce the formalism used to evaluate the influence of thermal noise in the schemes detailed in the previous section. First, we describe the notation, and then we formally introduce the considered physical interaction and how it is affected by the uncertainty of its position. We also introduce the logical systems and show how they can be used to minimise the effects of thermal noise.

\subsection{Notation}

Throughout the paper, we make use of stochastic and nonstochastic variables. To clarify the usage of those, a stochastic variable is written in Roman style, e.g., ``x'', while nonstochastic variables are in Italic style, e.g., ``\textit{x}''. We write $\text{x} \sim \{ x, p(x) \}$, if x is distributed with probability $p(x)$, and $x$ denotes a particular realization of the variable.

We also distinguish between different kinds of vectors. On the one hand, we consider ``spatial vectors'' which are 2- or 3-dimensional real vectors and refer to space positions. We write spatial vectors with the standard arrow notation, i.e., ``$\vec{x}\,$''. On the other hand, we denote any other kind of vector with a bold symbol, i.e., $\boldsymbol{x}$. We also use both notations to refer to a list of spatial vectors, i.e., $\vec{\boldsymbol{x}} = (\vec{x}_1,\dots,\vec{x}_K)$.

\subsection{Minimal size setting: two physical qubit systems}

First, we consider the minimal size case, two physical qubit systems, $A$ and $B$, each located at a certain position, $\vec{r}$ and $\vec{q}$, respectively. We assume an antiferromagnetic long-range Ising interaction between the qubits; i.e., the interaction Hamiltonian is given by
\begin{equation*}
    H = \mu(\vec{r},\vec{q}\,) \, Z^A Z^B
\end{equation*}
where the coupling strength depends on the distance between the two-qubit systems and it is given by $\mu(\vec{r},\vec{q}\,) = J \left|\vec{r} - \vec{q}\,\right|^{-\gamma}$, where $J$ is the coupling constant and $\gamma \in \mathbbm{N}$. By letting the qubits interact for a time $\Delta t = \frac{\pi}{4 \mu}$, one can implement our target entangling gate 
\begin{equation}
    U = e^{-\ti \frac{\pi}{4} ZZ},
\end{equation}
which can be, e.g., transformed into a CZ or a CNOT with the help of local operations. 

Uncertainty in the coupling strength leads to a noisy implementation of $U$ which we denote as the \textit{ZZ-damping channel}. If the coupling strength is distributed as $\upmu \sim \{ \mu, p(\mu) \}$, the evolution of an arbitrary two-qubit state $\rho$ is given by
\begin{equation*}
    \mathcal{U} (\rho) = \int e^{-\ti (\mu \Delta t) ZZ} \rho \, e^{\ti (\mu \Delta t) ZZ} \, p(\mu) \, \mathrm{d}\mu.
\end{equation*}
We evaluate the performance of the interaction with the so-called Choi fidelity of the channel $\mathcal{U}$ to the ideal gate $U$; i.e., the fidelity of the gate is given by $F = \bra{\Phi_U} \Phi_{\mathcal{U}} \ket{\Phi_U}$, where $\Phi_{\mathcal{U}} = \mathcal{U}\otimes \id (\proj{\Phi})$ is the Choi (mixed) state of $\mathcal{U}$ and $\ket{\Phi_U} = U \otimes \id \ket{\Phi}$ is the Choi (pure) state of the target gate $U$, where $\ket{\Phi} = \frac{1}{2} \sum_{i,j=0}^1 \ket{ij}\ket{ij}$. In the case of the \textit{ZZ}-damping channel, one can see that the fidelity is given by
\begin{equation}
    \label{eq:F0}
    F(t) = \left\langle \cos^2 \!\left( \tfrac{\pi}{4} - \upmu \, \Delta t \right) \right\rangle,
\end{equation}
see Appendix \ref{appendix:F} for derivation.

As $\mu$ depends on $\vec{r}$ and $\vec{q}$, thermal noise is a source of uncertainty. In particular, if $p(\vec{r}, \vec{q} \,)$ is the probability distribution of $\vec{\text{r}}$ and $\vec{\text{q}}$ then the probability distribution of $\upmu$ is given by
\begin{equation*}
    p(\mu) = \int \delta[\mu-\mu(\vec{r}, \vec{q}\,)] \, p(\vec{r}, \vec{q}\,) \, \mathrm{d}\vec{r} \, \mathrm{d}\vec{q},
\end{equation*}
where $\delta[x]$ is the Dirac delta function.

Note that if one is restricted to individual control, the most natural way to enhance the interaction between two qubits is by cooling down the physical systems, i.e., reducing uncertainty in their position. In the following section, we show how by using physical multiqubit systems to implement logical qubits, the fidelity can be enhanced without cooling down the system.

\subsection{Arbitrary size setting: two logical qubits}

We consider two modules $A$ and $B$ consisting of $N_A$ and $N_B$ spatially distributed physical qubit systems, respectively, i.e., $A = \{ A_i, \, \vec{r}_i \}_{i=1}^{N_A}$ and $B = \{ B_j, \, \vec{q}_j \}_{j=1}^{N_B}$, where physical qubit system $A_i(B_j)$ is located at the position $\vec{r}_i (\vec{q}_j)$. Like in the previous setting, we assume the qubits can be individually controlled but two-qubit gates cannot be implemented.

\subsubsection{Logical interaction}

The physical systems interact with a two-body long-range interaction given by
\begin{equation}
    \label{eq:Hzz}
    H_{\text{zz}} = H_{\text{zz}}^A + H_{\text{zz}}^B + H_{\text{zz}}^{AB},
\end{equation}
where
\begin{align*}
    & H_{\text{zz}}^A = \sum_{1\leq i < j \leq N_A} \mu(\vec{r}_i,\vec{r}_j) \, Z^A_i Z^A_j \\
    & H_{\text{zz}}^B = \sum_{1\leq i < j \leq N_B} \mu(\vec{q}_i,\vec{q}_j) \, Z^B_i Z^B_j \\
    & H_{\text{zz}}^{AB} = \sum_{\substack{1\leq i \leq N_A \\ 1 \leq j \leq N_B}} \mu(\vec{r}_i,\vec{q}_j) \, Z^A_i Z^B_j,
\end{align*}
where $H_{\text{zz}}^{A(B)}$ describes interactions within module $A(B)$, which we refer to as the \textit{self-interactions} of $A(B)$, and $H^{AB}$ describes interactions between physical systems in different modules.

Let us now prepare (encode) each module into a logical qubit subspace $span\{\ket{\bo{a}}, \ket{-\bo{a}}\}$ and $span\{\ket{\bo{b}}, \ket{-\bo{b}} \}$ by choosing the logical vectors $\bo{a}$ for $A$ and $\bo{b}$ for $B$. When the states of the modules are restricted to a logical subspace the Hamiltonian simplifies. On the one hand, the self-interaction terms just yield a global phase and can be ignored, i.e., $H_{\text{zz}}^A \ket{\pm \bo{a}} = f(\bo{a}) \ket{\pm \bo{a}}$ where $f(\bo{a}) = \sum_{i<j} a_i \, a_j \, \mu(\vec{r}_i,\vec{r}_j)$. On the other hand, $H^{AB}_{\text{zz}}$ is also diagonal in the logical subspace with eigenvalues 
\begin{equation*}
\begin{aligned}
    & H_{\text{\text{zz}}}^{AB} \ket{\pm \bo{a}, \pm \bo{b}} = \bar{\mu}^{\bo{ab}}(\vec{\bo{r}},\vec{\bo{q}}) \ket{\pm\bo{a}, \pm \bo{b}}, \\
    & H_{\text{\text{zz}}}^{AB} \ket{\pm \bo{a}, \mp \bo{b}} = - \bar{\mu}^{\bo{ab}}(\vec{\bo{r}},\vec{\bo{q}}) \ket{\pm\bo{a}, \mp \bo{b}}
\end{aligned}
\end{equation*}
where
\begin{equation*}
    \bar{\mu}^{\bo{a} \bo{b}}\! \left(\vec{\bo{r}}, \vec{\bo{q}} \right) = \!\! \sum_{\substack{1\leq i \leq N_A \\ 1 \leq j \leq N_B}} a_i \, b_j \, \mu(\vec{r}_i,\vec{q}_j),
\end{equation*}
is the coupling strength between the logical qubits, $\vec{\bo{r}} =  (\vec{r}_1,\dots, \vec{r}_{N_A})$ and $\vec{\bo{q}} =  (\vec{q}_1,\dots, \vec{q}_{N_B})$. Hence, for the logical qubits the Hamiltonian $H_{\text{zz}}$ simply reads
\begin{equation*}
    \bar{H} = \bar{\mu}^{\bo{a}\bo{b}} \! \left(\vec{\bo{r}},\vec{\bo{q}} \right) \, \bar{Z}^A \, \bar{Z}^B,
\end{equation*}
where $\bar{Z} = \proj{\bar{0}} - \proj{\bar{1}}$.

Note that the interaction between the logical qubits is also given by a \textit{ZZ} coupling, but its strength can be tuned by a suitable choice of the logical subspace, i.e., a suitable choice of the vectors $\bo{a}$ and $\bo{b}$.


\subsubsection{Logical gate fidelity}
\label{sec.logical.gate.fidelity}

Like in the minimal size setting, uncertainty in the position of the qubit systems leads to a noisy implementation of the two-logical-qubit gate. However, in contrast to the minimal size scenario, in this case, the effective coupling, and hence the gate fidelity with our target gate $U=e^{-\ti \frac{\pi}{4} ZZ}$, also depends on the logical subspaces $\bo{a}$ and $\bo{b}$,
\begin{equation}
\begin{aligned}
    \label{eq.Fab}
    F^{\bo{a}\bo{b}}(\Delta t) & = \left\langle \cos^2\!\left( \tfrac{\pi}{4} - \bar{\upmu}^{\bo{ab}} \, \Delta t \right) \right\rangle \\
    & = \int \!\! \cos^2 \!\left[ \tfrac{\pi}{4} - \bar{\mu}^{\bo{ab}}(\vec{\bo{r}}, \vec{\bo{q}}) \, \Delta t \right] p(\vec{\bo{r}}, \vec{\bo{q}})\, \mathrm{d}\vec{\bo{r}} \, \mathrm{d}\vec{\bo{q}}.
\end{aligned}
\end{equation}
Given a particular setting with a certain probability distribution for the position of the qubit systems, $p(\vec{\bo{r}},\vec{\bo{q}})$, our goal is to find the logical subspaces (given by $\bo{a}$ and $\bo{b}$) which maximize the fidelity and minimize the implementation time. In order to quantify the performance of a given setting, we compute its \textit{optimal infidelity curve} which corresponds to the optimal infidelity for every implementation time, i.e.,
\begin{equation}
    \label{eq.infidelitycurve}
    1- F^*(\Delta t) = 1 - \max_{\bo{a},\bo{b}} \, F^{\bo{a}\bo{b}} (\Delta t).
\end{equation}
In the following section, we analyze for different schemes how the fidelity can be enhanced by increasing the system size.


\begin{figure*}
    \centering
    \subfloat[]{\includegraphics[width=0.66\columnwidth]{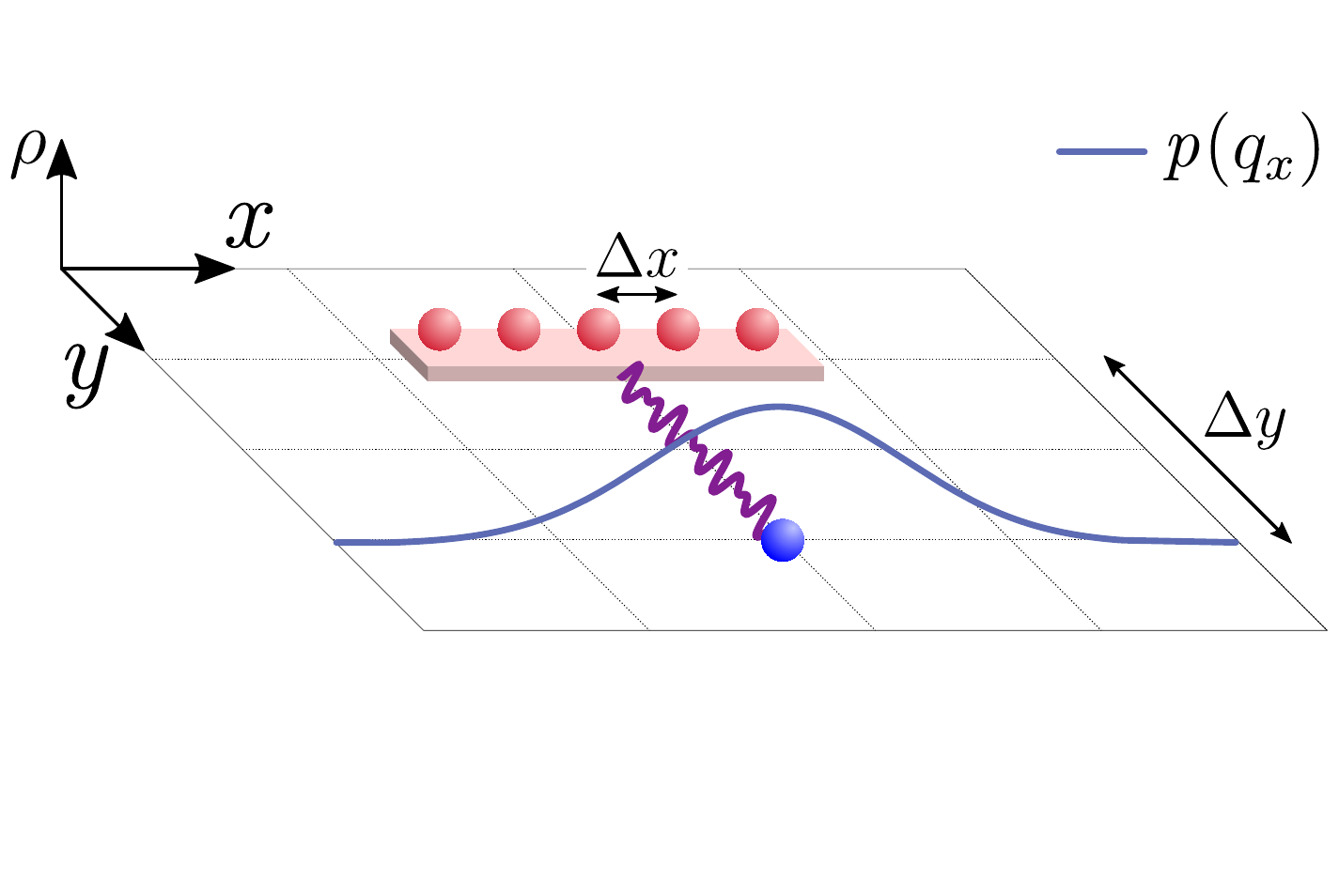}\label{Fig:frozen_1D}} \hspace{0.2cm}
    \subfloat[]{\includegraphics[width=0.66\columnwidth]{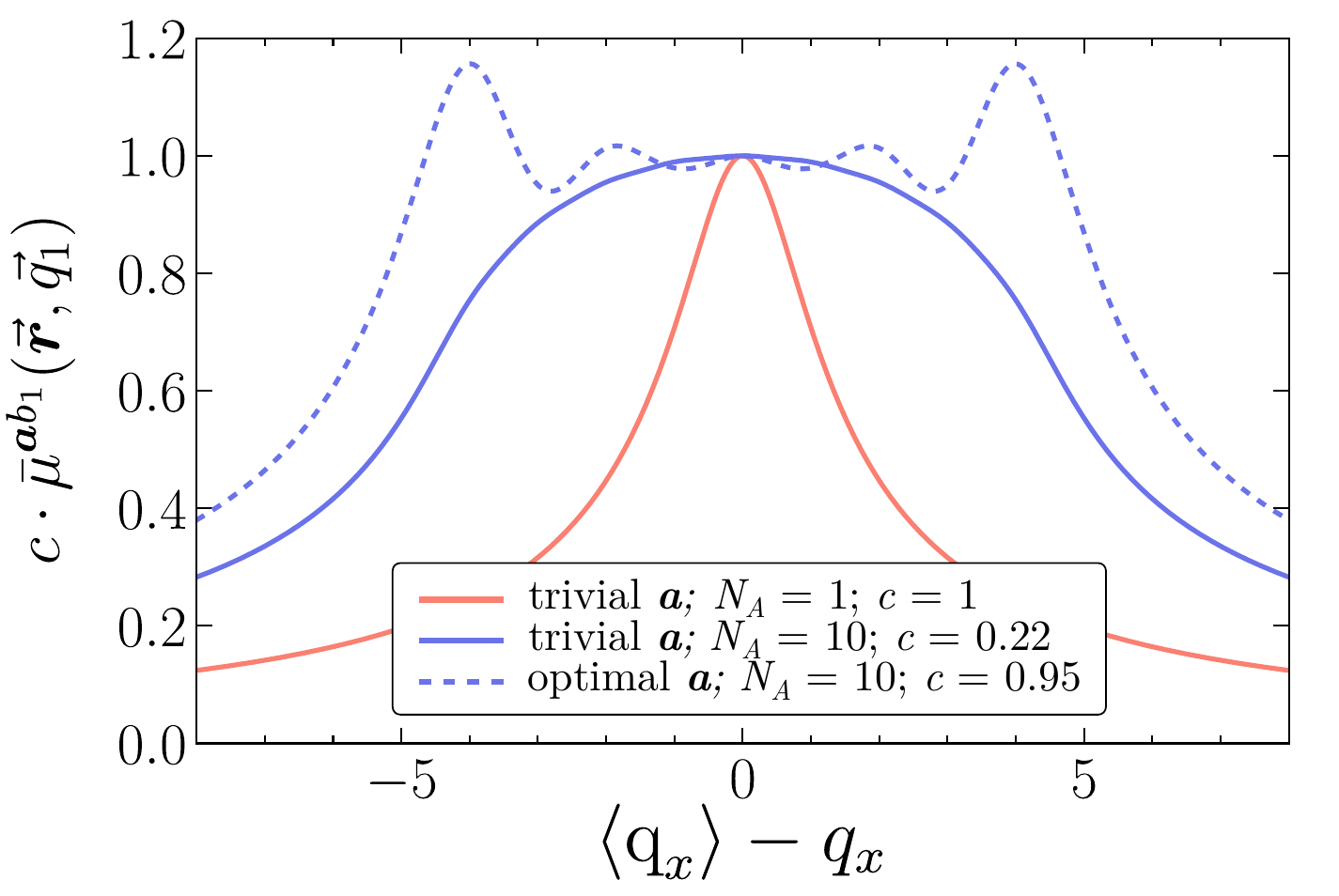} \label{plot:flat.f.lambda}} 
    \subfloat[]{\includegraphics[width=0.66\columnwidth]{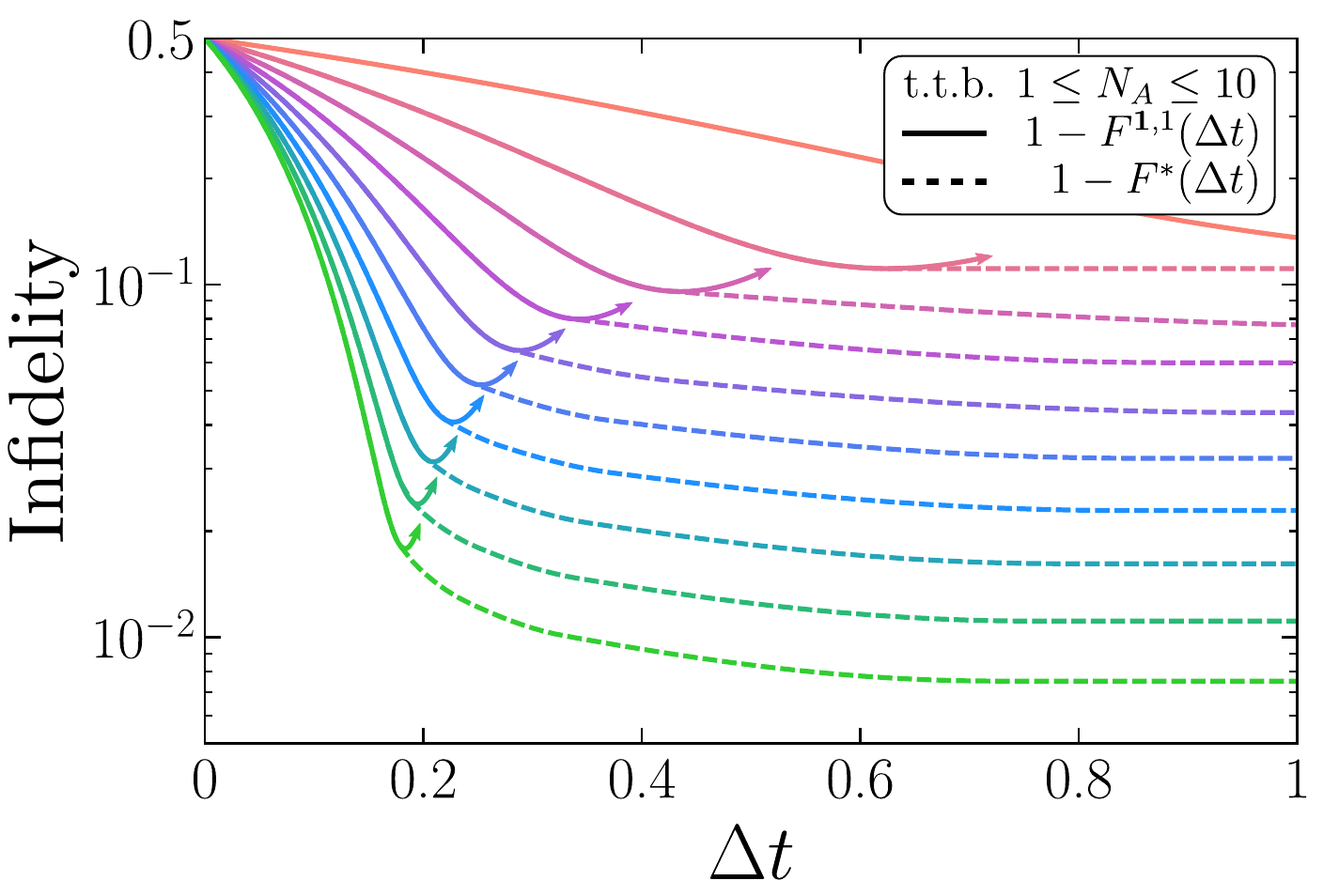} \label{plot:frozen_1D}} \\
    \subfloat[]{\includegraphics[width=0.66\columnwidth]{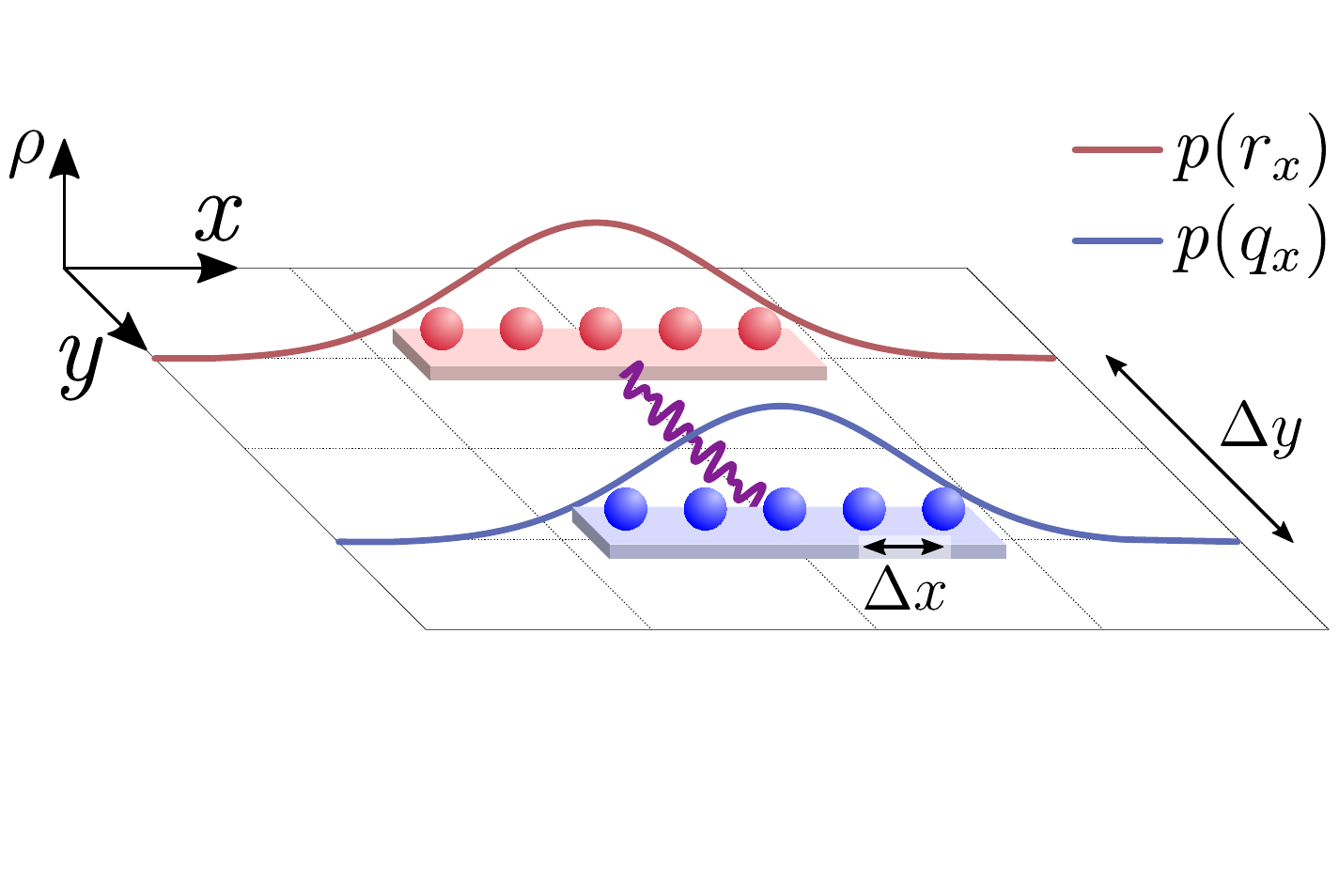}\label{Fig:collective_1D}} \hspace{0.2cm}
    \subfloat[]{\includegraphics[width=0.66\columnwidth]{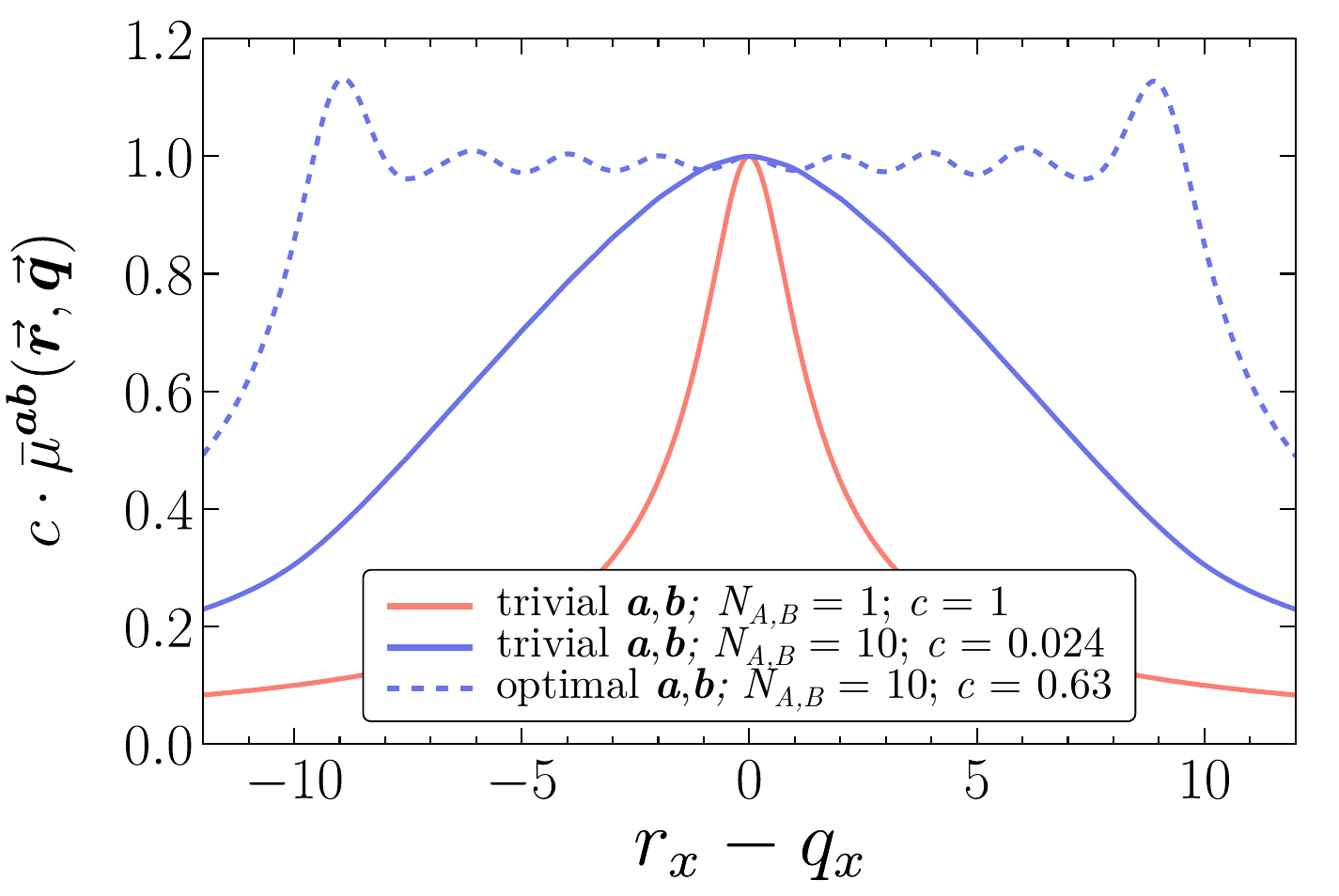} \label{plot:flat.c.lambda}} 
    \subfloat[]{\includegraphics[width=0.66\columnwidth]{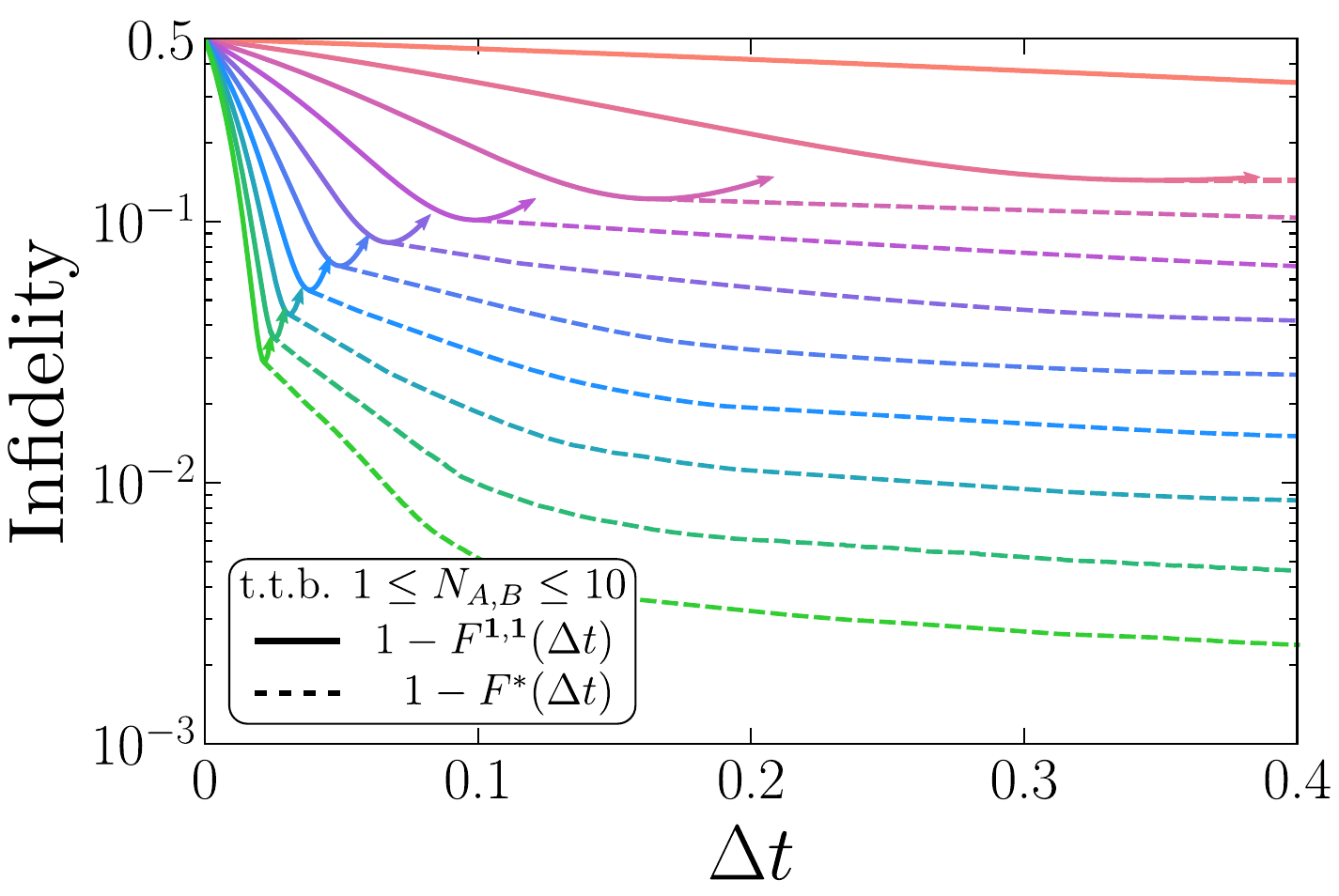} \label{plot:collective_1D}} 
    \caption{\label{fig:1} (a) Module $A$ is fixed and it consists of a 1D chain oriented along the $x$ axis, i.e., $\vec{r}_i = (i \Delta x, 0)$. Module $B$ consists of a single qubit system located at $\vec{\text{q}}_1 = (\text{q}_{1x}, \Delta y)$ where $\text{q}_{1x} \sim \mathcal{N}\left[(N_A-1)\Delta x/2, \sigma^2\right]$. (d) Each module, $A$ and $B$, consists of a 1D chain oriented along the $x$ axis, i.e., $\vec{\text{r}}_i = (i \Delta x + \text{r}_x, 0)$ and $\vec{\text{r}}_j = (j \Delta x + \text{r}_x, \Delta y)$ where $\text{r}_x$ and $\text{q}_x$ are i.i.d. $\sim \mathcal{N}\left[0,\sigma^2\right]$. (b, e) Logical coupling strength as a function of $\text{q}_x$ and $\text{r}_x-\text{q}_x$ for the settings described in (a) and in (d), respectively. Each curve is multiplied by a constant $c$ such that the difference in shape can be appreciated. Solid lines correspond to the trivial encoding, i.e., $\bo{a} = \bo{b} = \bo{1}$, while dotted lines to the optimal encoding for $t = 0.9$ and $0.5$, respectively. (c, f) Infidelity of the implementation of $U= e^{-\ti \frac{\pi}{4}ZZ}$ as a function of the implementation time $\Delta t$ for different system sizes for the setting described in (a) and in (d), respectively, with $J = 1$, $\gamma=1$, $\Delta x = \Delta y = 1$ and $\sigma=3$. ``t.t.b.'' means ``top to bottom''.}
\end{figure*}

\section{Cold mediating system}
\label{sec.Cold.mediating.system}

First, we look at the scheme detailed in Sec.~\ref{sec.Interactions.mediator.system} where a fully controllable cold system is used to mediate an effective interaction between two-qubit systems affected by thermal noise. If we consider the gate sequence shown in Eq.~\eqref{eq:mediating}, the fidelity of the mediated interaction is given by $F_{12} = F_{01} F_{02}$ where $F_{0i}$ is the fidelity of a control gate $U$ between the $i$th qubit and the auxiliary system (see Appendix.~\ref{Appendix:F1F2} for derivation). Therefore, for our purposes, it suffices to analyze the implementation of $U = e^{-\ti \frac{\pi}{4} ZZ}$ between a logical and a hot and a single qubit affected by position noise.

The scheme we consider is inspired by 1D traps where particles are strongly bounded in the $y$ direction but weakly in the $x$ direction. Module $A$ contains $N_A$ perfectly trapped physical qubit systems, meaning their positions are well defined. In turn, $B$ consists of a single physical qubit system, $N_B = 1$, with a noisy position, i.e., $\vec{\textbf{q}} = \vec{\text{q}}_1$ and $\bo{b} = (b_1)$. In this case, the logical coupling strength is given by $\bar{\mu}^{\bo{a}b_1}(\vec{\bo{r}},\vec{\text{q}}_1) = b_1 \sum_{i=1}^{N_A} a_i \, \mu(\vec{r}_i , \vec{\text{q}}_1)$. We consider the setting shown in Fig.~\ref{Fig:frozen_1D}. Module $A$ is a 1D chain aligned with the $x$ axis. System $B$ is at a fixed distance $\Delta y$ of the chain, but its position along the $x$ is a normally distributed random variable, i.e., $\vec{\text{q}}_1 = (\text{q}_{1x}, \Delta y)$ where $\text{q}_{1x} \sim \mathcal{N}\!\left[ 0, \, \sigma^2\right]$.

As shown in the previous section, the logical coupling strength $\bar{\mu}^{\bo{a}b_1}(\vec{\bo{r}},\vec{q}_1)$ can be modified by changing the logical vector $\bo{a}$. Ideally, we would like to establish a logical coupling which is independent of the position of the physical qubit system $\vec{q}_1$, making the interaction insensitive to thermal fluctuations. However, this is not possible due to the inherent form of the physical qubit-qubit interaction. Nevertheless, we can establish a $\bar{\mu}^{\bo{a}b_1} (\vec{\bo{r}},\vec{q}_1)$ that approximates a constant function within the region of space where qubit $B$ is most likely to be found. In Fig.~\ref{plot:flat.f.lambda} we plot the logical coupling strength as a function of $q_{1x}$. Observe  how with the trivial encoding increasing the size of $A$, i.e., $\bo{a} = \bo{1} = (1, \dots, 1)$, $\bar{\mu}^{\bo{a}b_1}(\vec{\bo{r}},\vec{q}_1)$ becomes flatter for $\text{q}_{1x} \in [-\sigma, \sigma]$. We can increase this effect even further by optimising the logical subspace, see in Appendix. \ref{appendix:optimal.logical.subspaces} the optimal $\bo{a}$ for different system sizes. Note that the optimal logical subspace already amplifies the interaction. Still, the maximum coupling strength is obtained with the trivial encoding, i.e., $\max_{\bo{a},b_1} \bar{\mu}^{\bo{a}b_1} = \bar{\mu}^{\bo{1}1}$.

In Fig.~\ref{plot:frozen_1D} we plot the infidelity curve Eq.~\eqref{eq.infidelitycurve} for different values of $N_A$, i.e., for each implementation time $\Delta t$ we show the best infidelity we can reach by establishing a logical subspace $\bo{a}$ for all $\Delta t$. If $\Delta t$ is small, the fidelity is given by $F^{\bo{a}\bo{b}} (\Delta t) = \frac{1}{2} + \left\langle \bar{\upmu}^{\bo{ab}} \right\rangle \Delta t$, and hence, in this regime, the trivial encoding, i.e., $\bo{a} = (1,\dots,1)$, is optimal because it maximizes $\bar{\upmu}^{\bo{ab}}$. At a certain value of $\Delta t$, the trivial encoding is no longer optimal, as it is better to establish a weaker but flatter coupling function $\bar{\mu}^{\bo{a}b_1}(\vec{\bo{r}},\vec{q}_1)$. This trade-off gradually changes with $\Delta t$ until eventually the optimal shape $\bar{\mu}^{\bo{a}b_1}(\vec{\bo{r}},\vec{q}_1)$ can be established, as the drop in interaction strength can be compensated by the longer interaction time. Then the infidelity saturates because one always can ``slow down'' the interaction by scaling the optimal logical encoding, i.e., $F^{\bo{ab}}(\Delta t) = F^{\bo{a}' \bo{b}}( \Delta t / c )$ where $\bo{a}' = c \, \bo{a}$. In the figure, we show the best fidelity that can be achieved with the trivial encoding (solid lines) and the enhancement (dashed liens) that an optimisation of $\bo{a}$ provides. Observe that the larger the size of $A$ the bigger the enhancement in terms of fidelity and implementation time. To compute the optimal infidelity curve, we discretize $\Delta t$, and for each $\Delta t$ we used the Wolfram Mathematica function {\ttfamily NMaximize} to find the optimal logical encoding. To avoid local minima we used the solution of the previous time step as a starting point for the next optimization. In Appendix \ref{sec.Cold.mediating.system.2D} we show the optimal infidelity for a 2D example. This behavior is observed in all examples computed in the article and the same discussion applies to all of them.


\section{Collective position noise}
\label{sec.Collective.position.noise}

In this section, we analyze the scheme introduced in Sec.~\ref{sec.Independent.classical.modules}. We consider modules $A$ and $B$ to be affected by collective noise. The position of the physical qubit systems is well defined within the modules, whose ``center of mass'' position is subject to noise. Formally, this means that we can parametrize the position of the physical qubit systems of each module as
\begin{align*}
    \vec{\textbf{r}} & = \vec{\bo{r}}^{\,0} + (\vec{\text{r}}, \vec{\text{r}}, \dots, \vec{\text{r}}\,), \\
    \vec{\textbf{q}} & = \vec{\bo{q}}^{\,0} + (\vec{\text{q}}, \vec{\text{q}}, \dots, \vec{\text{q}}\,),
\end{align*}
where $\vec{\text{r}}$ and $\vec{\text{q}}$ are two independent stochastic variables, and $\vec{\bo{r}}^{\,0}$ and $\vec{\bo{q}}^{\,0}$ are the positions of the physical qubit systems in the module reference frame, which are fixed.

Here we consider each module to be arranged in a 1D chain along the $x$ axis; see Fig.~\ref{Fig:collective_1D}. Similarly to the example considered in Sec.~\ref{sec.Cold.mediating.system}, their position on the $y$ axis is fixed but not on the $x$ axis, i.e., $\vec{\text{r}} = (\text{r}_x, r_y)$ and $\vec{\text{q}} = (\text{q}_x, q_y)$, while $\text{r}_x$ and $\text{q}_x$ are independent and identically distributed (i.i.d.) $\sim \mathcal{N}[0,\sigma^2]$. In Fig.~\ref{plot:flat.c.lambda} we plot the logical coupling strength as a function of the distance between the two modules, i.e., $\bar{\mu}^{\bo{ab}}(\text{r}_x - \text{q}_x)$. We obtain similar behavior as in the previous setting. In this case, the optimal subspaces lead to an approximate logical coupling strength for a significantly larger region, as the total number of physical qubit systems (and hence tunable parameters) is larger than the one in Fig.~\ref{plot:flat.f.lambda}. In Fig.~\ref{plot:collective_1D}, we show the optimal infidelity curve with our target gate $U$ for different system sizes. The results are similar to the ones obtained in Sec.~\ref{sec.Cold.mediating.system}. However, in this case, for each curve we increase the size of both modules, leading to a quadratic enhancement in the coupling strength.

Similar results are obtained when considering a 2D example; see Appendix \ref{sec.collective.position.noise.2D} for details.


\section{Independent noise}
\label{sec.Independent.noise}

We now consider the model detailed in Sec.~\ref{sec.Independent.physical.qubits}, where the position of each physical qubit system is an independent stochastic discrete variable, i.e., $p(\vec{\bo{r}}, \vec{\bo{q}}) = p(\vec{r}_1) \, p(\vec{q}_1) \cdots p(\vec{r}_{N_A}) \, p(\vec{q}_{N_B})$. In this case, the fidelity is given by
\begin{equation*}
\begin{aligned}
    F^{\bo{ab}}(t) = \sum_{\vec{\bo{r}},\vec{\bo{q}}} p(\vec{\bo{r}},\vec{\bo{q}}) \cos^2\!\left[ \tfrac{\pi}{4} - \bar{\mu}^{\bo{ab}}(\vec{\bo{r}}, \vec{\bo{q}}) \, \Delta t \right].
\end{aligned}
\end{equation*}
Note that in this case, if each physical qubit system can be in $\kappa$ different positions the total number of configurations and possible couplings $\bar{\mu}^{\bo{ab}}(\vec{\textbf{r}}, \vec{\textbf{q}})$ is given by $\kappa^{N_A+N_B}$. Unlike in the previous examples, in this case, increasing the size of the modules $A$ and $B$ the sample space of the logical coupling strength $\bar{\mu}$ also increases. However, we show that this does not prevent our setting from providing improvements in fidelity.

\begin{figure}
    \centering
    \subfloat[]{\includegraphics[width=0.75\columnwidth]{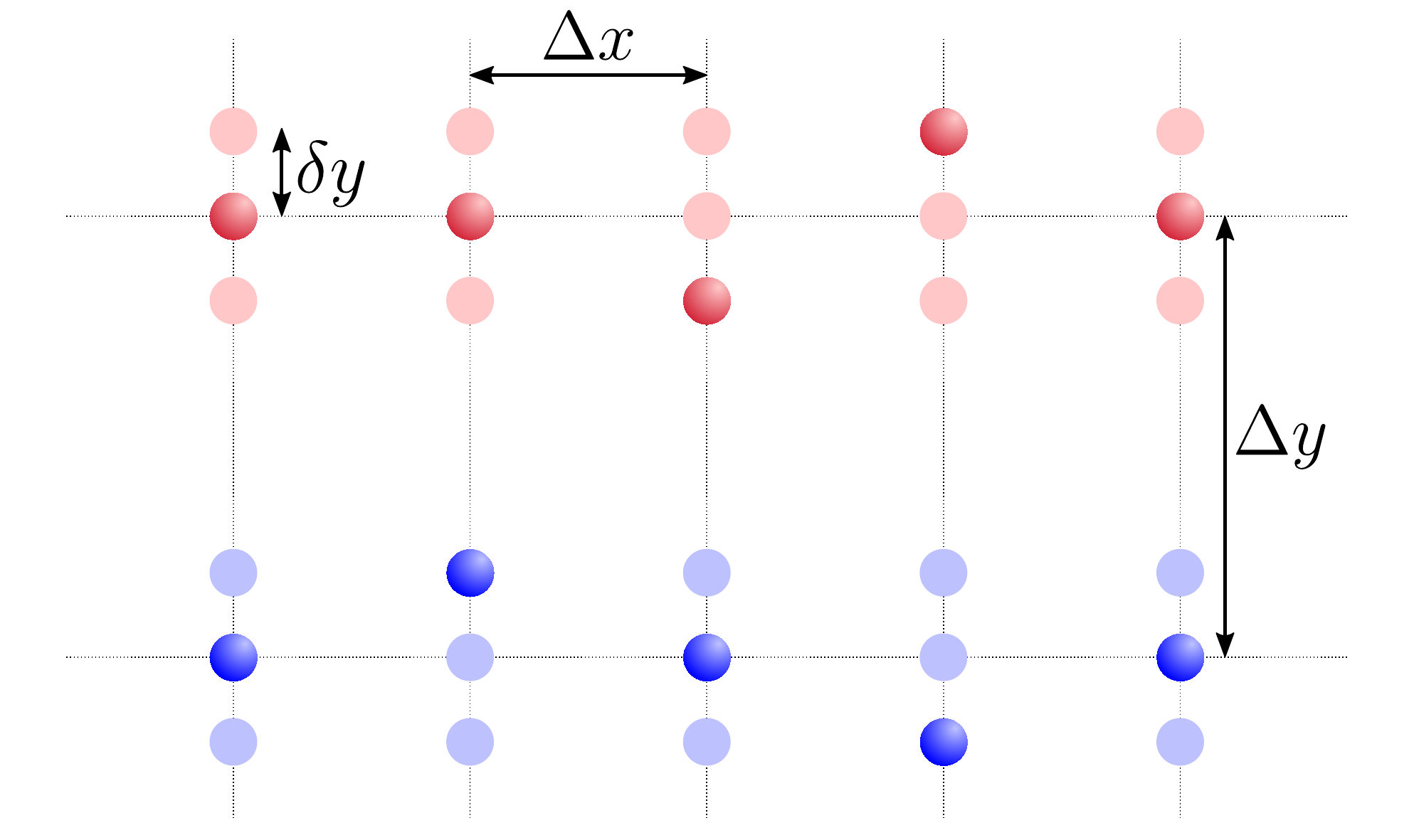}\label{fig:independent:cartoon}} \\
    \subfloat[]{\includegraphics[width=0.735\columnwidth]{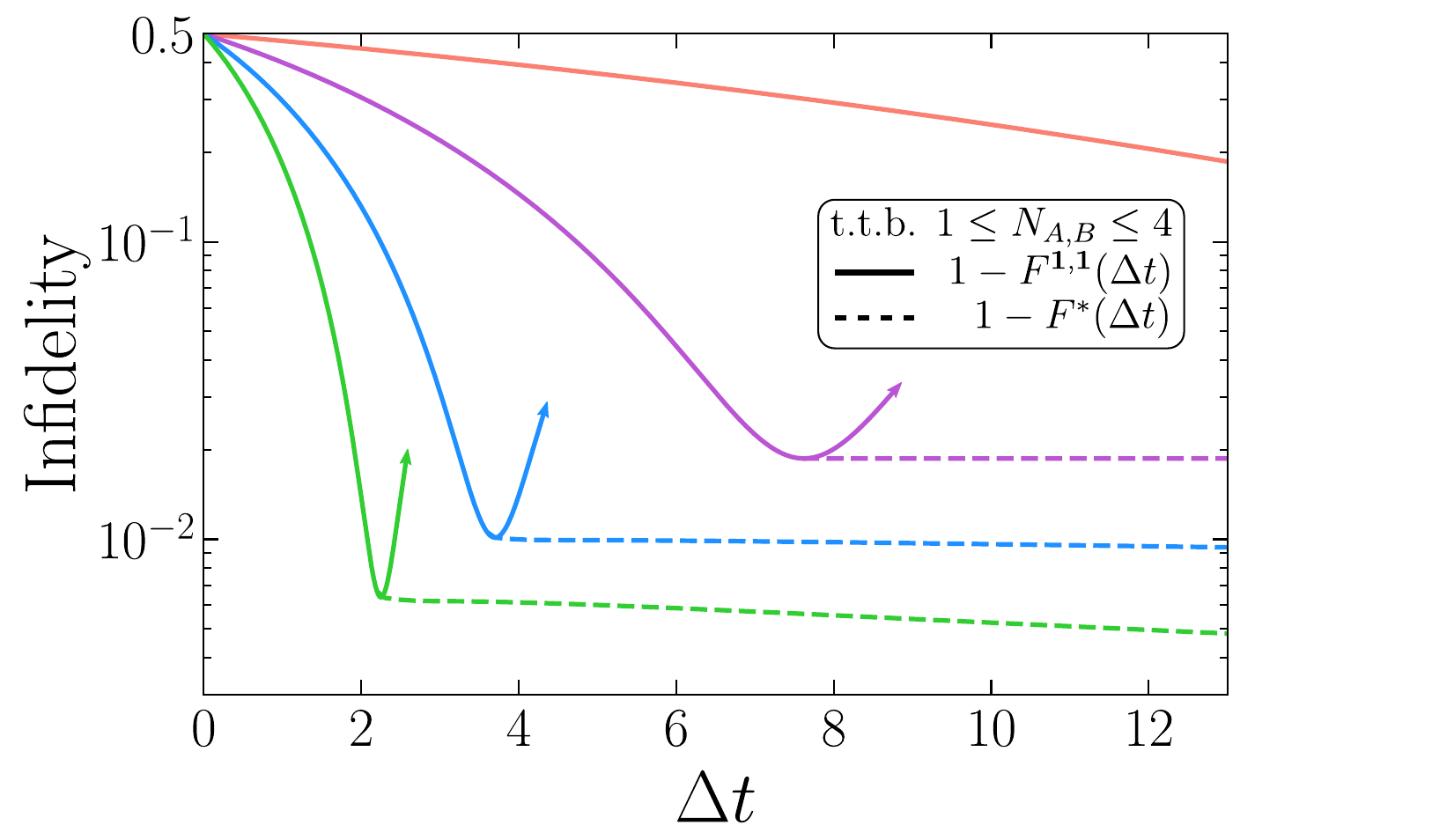}\label{fig:independent:plot}}
    \caption{\label{fig:independent} (a) The setting consists of two 1D chains oriented along the $x$ axis, where each system $A$ and $B$ corresponds to one of the chains. The positions of qubits $A_i$ and $B_j$ are given by $\vec{r}_i = (i \Delta x, \Delta y + \text{r}_{iy})$ and $\vec{q}_j = (j \Delta x, \text{q}_{yj})$, respectively, where $\text{r}_{xi}$ and $\text{q}_{yj} \in \{ -\delta y, 0, \delta y\}$, with $p(\pm\delta y) = 1/4$ and $p(0) = 1/2$. (b) Optimal infidelity of the implementation of $U= e^{-\ti \frac{\pi}{4}ZZ}$ as a function of the implementation time $\Delta t$ for different systems sizes for the setting described in (a), with $\Delta x = 2$, $\Delta y = 4$, $\delta y = 1$, $\gamma = 1$, and $J=1$.}
\end{figure}

In this case, we again consider each module as a 1D chain oriented in the $x$ axis. However, in order to avoid spatial overlap between the physical qubit systems while having significantly large position fluctuations, in this case, we assume fluctuations along the $y$ axis; see Fig.~\ref{fig:independent:cartoon}. Figure~\ref{fig:independent:plot} shows that the optimal infidelity is significantly improved for any value of $\Delta t$ when increasing the size of $A$ and $B$. However, in this case, due to computational limitations, we are restricted to considering small system sizes. As a result, we are unable to observe a notable improvement in fidelity from optimizing the encoding.

\section{Position quantization}
\label{sec.position.quantitzation}

In this last section, we look at fully quantized scenarios where the positions of the physical systems, $\vec{\bo{r}}$ and $\vec{\bo{q}}$, are described by quantum operators. In general, knowing the mechanical properties of the physical systems and the trapping potential used to confine them one can define the ``mechanical'' Hamiltonian $H_m$ that governs their motion. This Hamiltonian can be diagonalized
\begin{equation}
    \label{eq:Hamiltonian:position}
    H_{\text{m}} = \sum_{k=0}^{\infty} E_k \proj{E_k},
\end{equation}
to define the mechanical eigenmodes
\begin{equation*}
    \ket{E_k} = \int \Phi_k(\vec{\bo{r}}, \vec{\bo{q}}) \ket{\vec{\bo{r}}, \vec{\bo{q}}} \mathrm{d}\vec{\bo{r}} \, \mathrm{d}\vec{\bo{q}},
\end{equation*}
associated to the wave function $\Phi_k(\vec{\bo{r}}, \vec{\bo{q}})$ and energy $E_k$. The full Hamiltonian is now given by the sum of the two terms, 
\begin{equation}
    \label{eq:TOTAL:Hamiltonian:position}
    H = \mathbbm{1}_\text{Q} \otimes H_{\text{m}} + H_{\text{zz}},
\end{equation}
where $H_{\text{m}}$ does not act on the internal (qubit) degrees of freedom, and $H_{\text{zz}}$ is still given by Eq.~\eqref{eq:Hzz} but with the positions $\vec{r}_i$ and $\vec{q}_j$ which are operators [e.g., $ H_{\text{zz}}^A = \sum_{i, j}  Z^A_i Z^A_j \otimes \mu(\vec{r}_i,\vec{r}_j)$].

\begin{figure*}
    \centering
    \subfloat[]{\includegraphics[width=0.66\columnwidth]{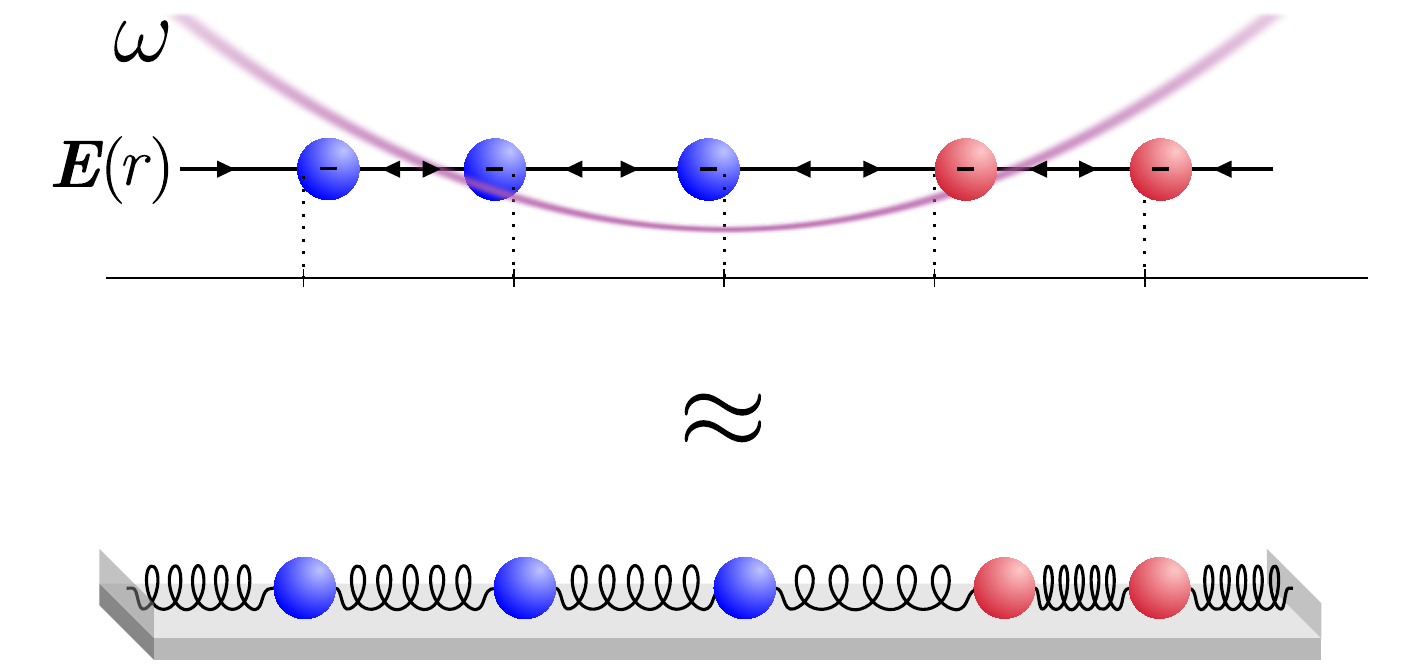} \label{fig:cartoon_1trap}} 
    \subfloat[]{\includegraphics[width=0.66\columnwidth]{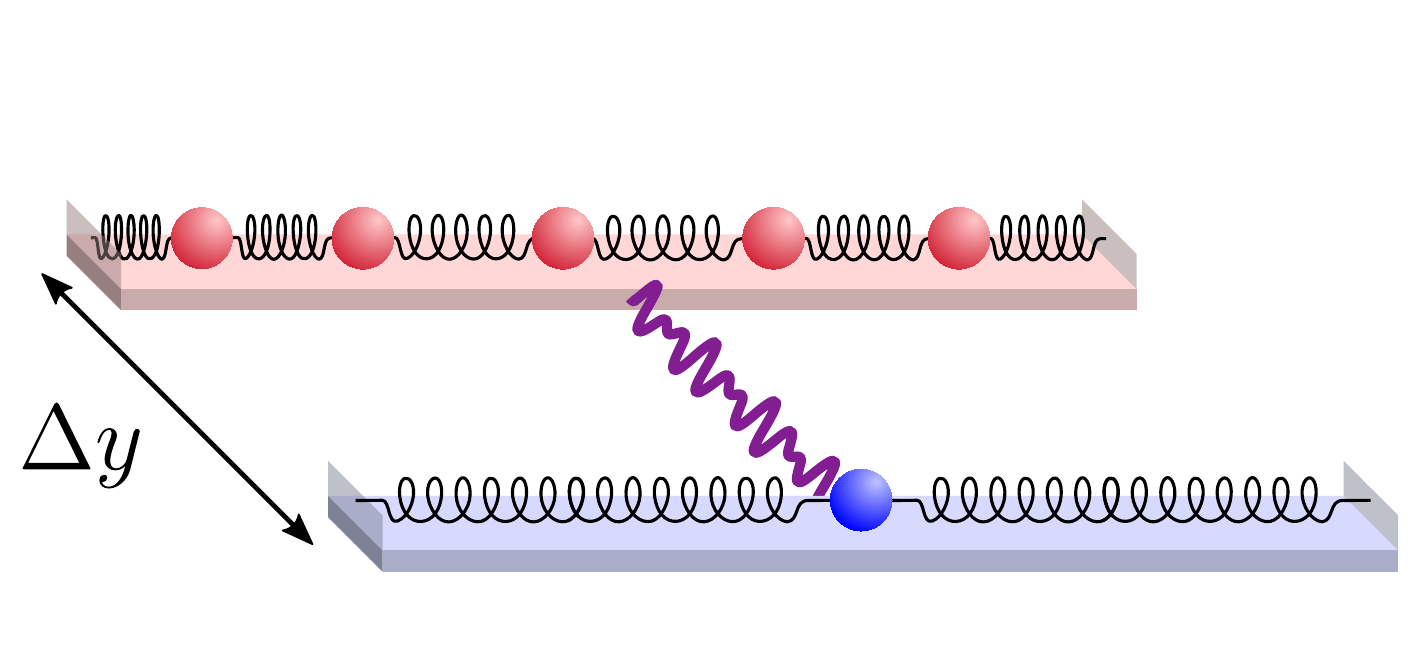} \label{fig:cartoon_cold_trap}}
    \subfloat[]{\includegraphics[width=0.66\columnwidth]{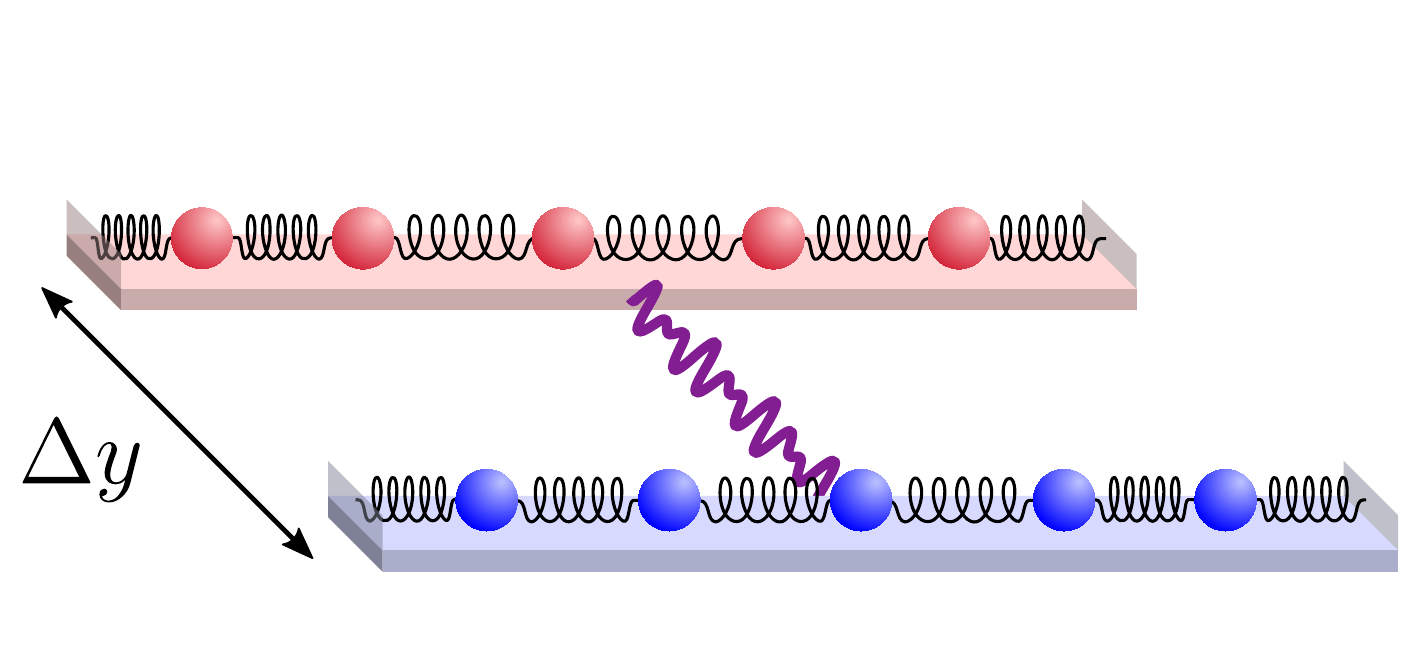} \label{fig:cartoon_2traps}} \\
    \subfloat[]{\includegraphics[width=0.66\columnwidth]{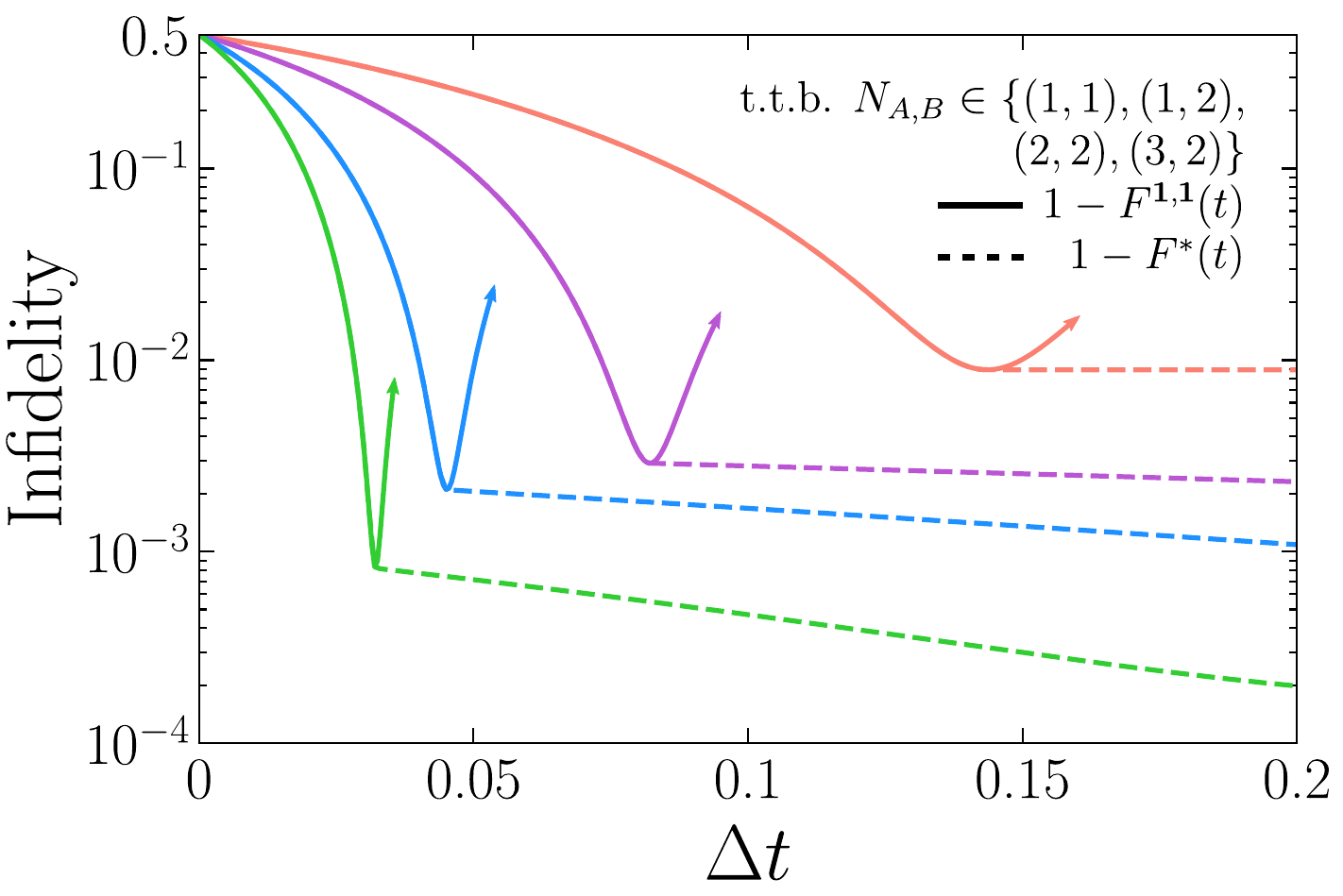} \label{fig:1trap}} 
    \subfloat[]{\includegraphics[width=0.66\columnwidth]{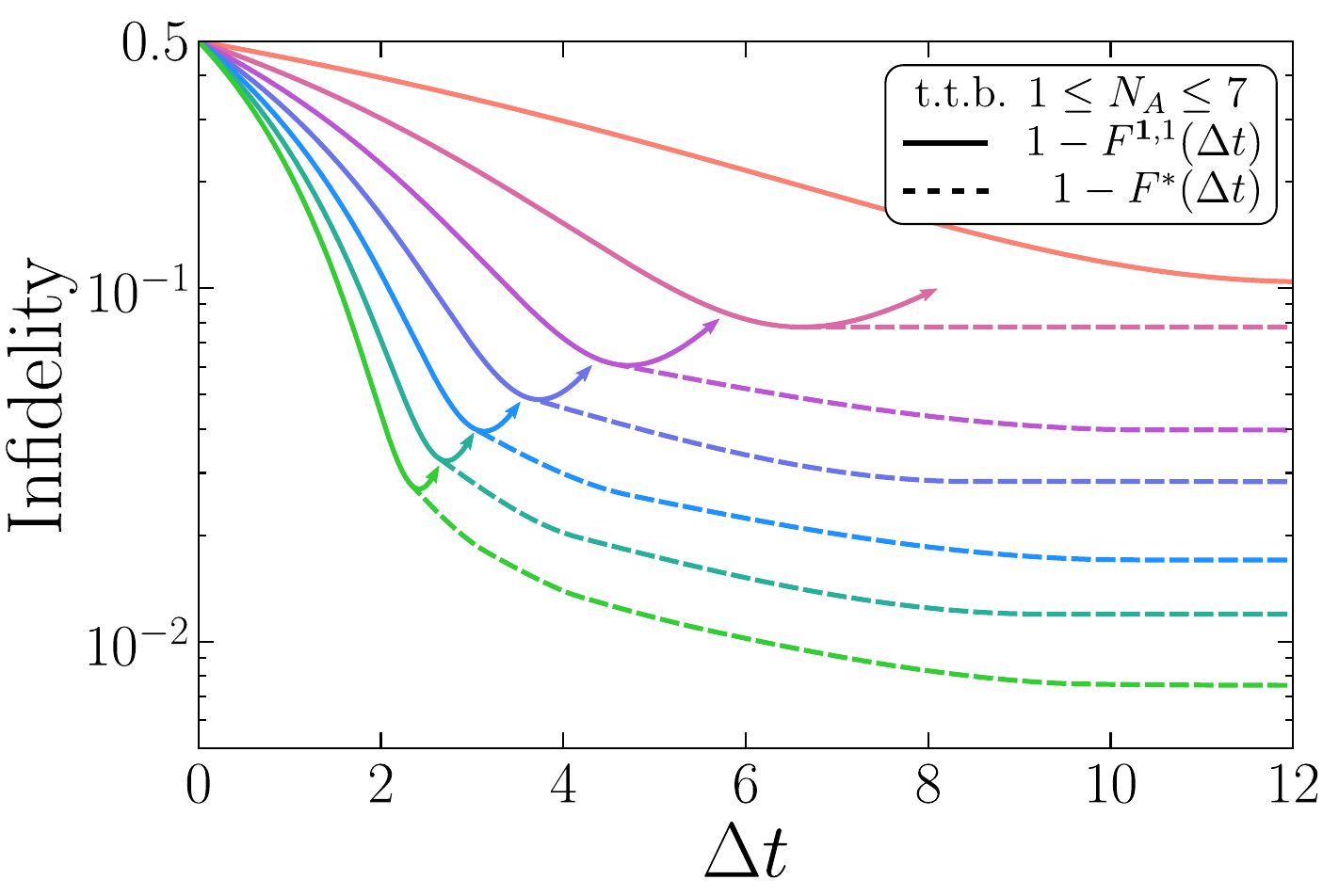} \label{fig:cold_trap}}
    \subfloat[]{\includegraphics[width=0.66\columnwidth]{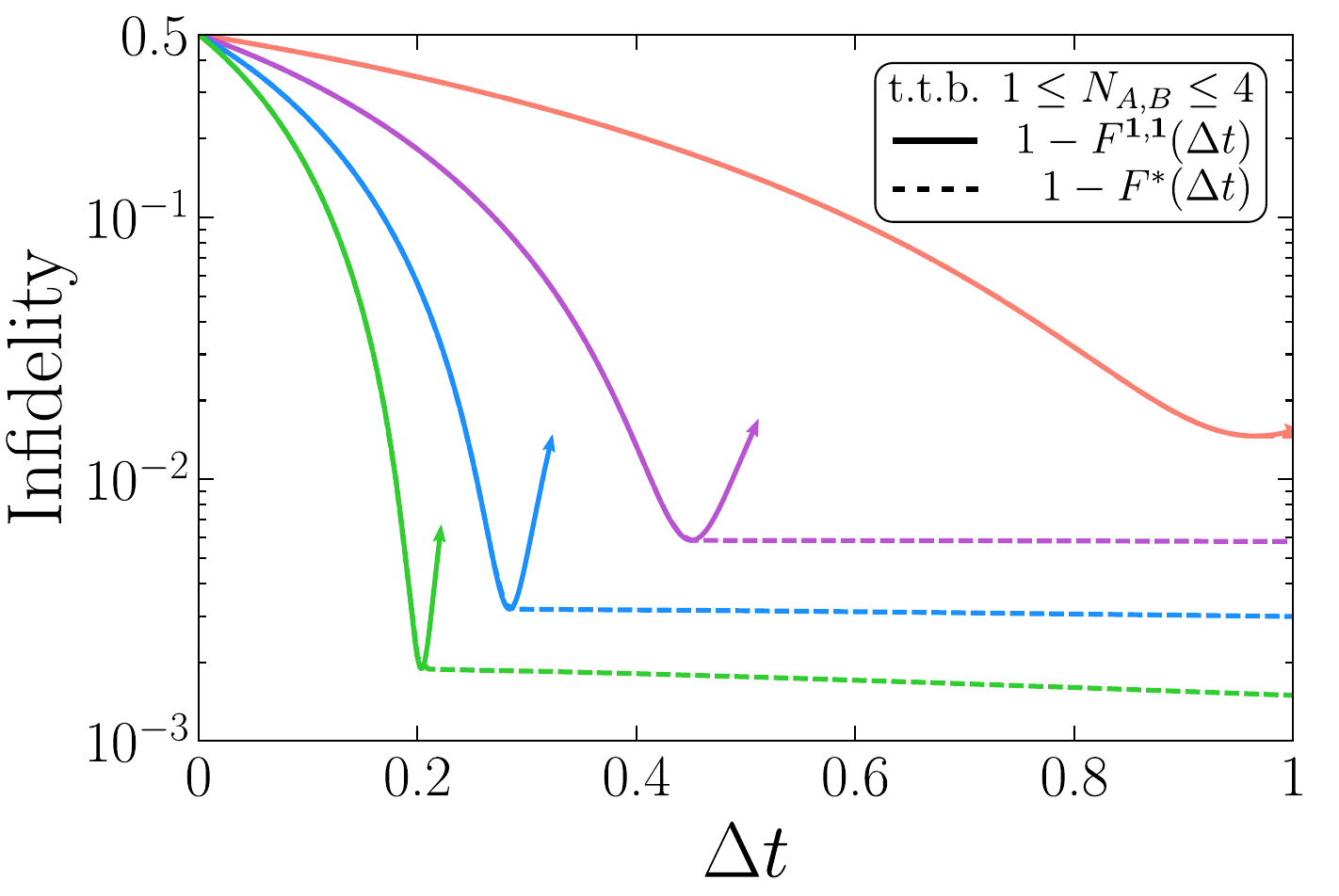} \label{fig:2traps}}
    \caption{\label{fig:Paul:traps} (a,b,c) Different settings consisting of 1D Paul traps. (d,e,f) Optimal infidelity of the implementation of $U= e^{-\ti \frac{\pi}{4}ZZ}$ as a function of the implementation time $\Delta $ for different system sizes, $N_A$ and $N_B$, with $\gamma = 3$ and $J=1$ for each of the settings. ``t.t.b.'' means ``top to bottom''. (a,d) A 1D chain of ions in a harmonic potential with a Coulomb interaction can be written in terms of the normal modes. System $A$ consists of the first half of the chain and system $B$ of the rest of the ions, with $\omega = 1$, $L = 4.78$, $T = 1.3$, and $\epsilon = 0.07$. (b,e) Systems $A$ and $B$ are in parallel independent traps with $N_B = 1$, $\omega_A = 100 \, \omega_B = 1$, $\Delta y = 20$, $L = 15.97$, $T = 0.1$, and $\epsilon = 0.05$. (c,f) Systems $A$ and $B$ are in parallel independent traps with $N_A = N_B$, $\omega_A = 3 \, \omega_B = 1$, $L_A = L_B = 8.31$, $\Delta y = 2$, $T = 0.2$, and $\epsilon = 0.01$. }
\end{figure*}

When we restrict $A$ and $B$ into logical subspaces and use the basis $\{\ket{E_k}\}$ for the mechanical dof, we can write the different terms of $\bar{H}_{\text{zz}}$ as
\begin{equation}
\begin{aligned}
    \label{eq:logical:Hamiltonian:position}
    \bar{H}_{\text{zz}}^A + \bar{H}_{\text{zz}}^B & = \sum_{k,l = 0}^{\infty} \left[ \left( \bar{\mu}^{\bo{a}}_{kl} + \bar{\mu}^{\bo{b}}_{kl} \right) \ketbra{E_k}{E_l} \; \right], \\
    \bar{H}_{\text{zz}}^{AB} & = \sum_{k,l = 0}^{\infty} \bar{\mu}^{\bo{a}\bo{b}}_{kl} \, \bar{Z}^A \, \bar{Z}^B \otimes \ketbra{E_k}{E_l}, \\
\end{aligned}
\end{equation}
where
\begin{align*}
    \bar{\mu}_{kl}^{\bo{a}} & = \sum_{1 \leq i < j \leq N_A} a_i \, a_j \bra{E_k} \mu(\vec{r}_i, \vec{r}_j) \ket{E_l}, \\
    \bar{\mu}_{kl}^{\bo{b}} & = \sum_{1 \leq i < j \leq N_B} b_i \, b_j \bra{E_k} \mu(\vec{q}_i, \vec{q}_j) \ket{E_l}, \\
    \bar{\mu}_{kl}^{\bo{ab}} & = \sum_{\substack{1\leq i \leq N_A \\ 1 \leq j \leq N_B}} a_i \, b_j \bra{E_k} \mu(\vec{r}_i, \vec{q}_j) \ket{E_l}.
\end{align*}

Note that the self-interaction terms act trivially on the logical qubits; nevertheless, they induce transitions between mechanical eigenmodes contributing to the noise.

For simplicity, we assume the system is initialized in a state of the form $\rho_{\text{m}} \otimes \bar{\rho}$, where $\bar{\rho}$ is an arbitrary state of the two logical qubits and
\begin{equation*}
    \rho_{\text{m}} = \sum_{k=0}^\infty p_k \proj{E_k},
\end{equation*}
is an arbitrary density operator commuting with $H_\text{m}$. Then the evolution of the logical qubits is given by
\begin{equation}
\begin{aligned}
\label{eq:Uquantized}
    \mathcal{U}(\bar{\rho}) & = \text{tr}_{\text{m}} \! \left[ \, e^{-\ti \Delta t \, H} \left( \rho_{\text{m}} \otimes \bar{\rho} \right) e^{\ti \Delta t \,  H} \, \right].
\end{aligned}
\end{equation}

\subsection{Non-degenerate case}
\label{sec.traps}

First, we discuss the case where $H_{\text{m}}$ is non-degenerate, and assume that there is a clear scale separation between the mechanical and the logical qubit interaction energies, i.e., $|E_k-E_l| \gg \left|\bar{\mu}^{\bo{a} \bo{b}}\right|$. We can then go to the interaction picture $H' = \ti \, \Dot{W} W^\dagger + W H W^\dagger$ where $W = e^{-\ti H_{\text{m}} t}$ and perform the rotating wave approximation. This results in neglecting the hopping between mechanical eigenstates induced by the interaction between logical qubits, i.e., the terms $\bar{\mu}_{kl}^{\bo{ab}}$ with $k\neq l$, and leads to the following Hamiltonian,
\begin{equation*}
    H' = \sum_{k=0}^{\infty} \left[ \Big( \bar{\mu}^{\bo{a}}_k + \bar{\mu}^{\bo{b}}_k + \bar{\mu}^{\bo{a}\bo{b}}_k \, \bar{Z}^A \bar{Z}^B \Big) \otimes \ketbra{E_k}{E_k} \; \right],
\end{equation*}
where $\bar{\mu}_k = \bar{\mu}_{kk}$.

Note that if we let the system evolve the logical qubits interact with a logical coupling strength that is a discrete stochastic variable distributed as $\bar{\upmu}^{\bo{ab}} \sim \{ \bar{\mu}_k^{\bo{ab}}, p_k\}$; i.e., Eq.~\eqref{eq:Uquantized} becomes
\begin{equation*}
    \mathcal{U}(\bar{\rho}) = \sum_{k=1}^\infty p_k \, e^{-\ti (\bar{\mu}_k^{\bo{ab}} \, \Delta t ) ZZ} \, \bar{\rho} \; e^{\ti (\bar{\mu}_k^{\bo{ab}} \, \Delta t) ZZ}.
\end{equation*}
Therefore, the fidelity is given by
\begin{equation*}
    F^{\bo{ab}}(\Delta t) = \sum_{k=0}^\infty p_k \cos^2\!\left( \tfrac{\pi}{4} - \bar{\mu}^{\bo{ab}}_{k} \, \Delta t \right).
\end{equation*}
Observe that in this case, the logical qubits are also insensitive to the self-interactions.
 
\subsubsection{Hot 1D Paul trap}

We now consider a particular example where the physical qubit systems are ions in a 1D Paul trap. The ions are assumed to be strongly bounded in the $y$ and $z$ directions but weakly trapped in a harmonic potential along the $x$ axis. The ions also interact with each other with a Coulomb interaction that leads to collective dynamics within the trap. A treatment of the mechanical motion of the ions, which allows us to derive the mechanical eigenstates, can be found in Ref. \cite{hot_james1997quantum} (see also Appendix \ref{appendix:trapmodes}). In addition, we assume the system is in contact with a thermal bath at temperature $T$, and $\rho_{\text{m}}$ is given by a thermal state, i.e.,
\begin{equation*}
    p_k = \frac{e^{-E_k/T}}{\sum_{l=0}^\xi e^{-E_l/T}},
\end{equation*}
where ideally one should consider $\xi \to \infty$. However, due to computation limitations, we are forced to consider a finite $\xi$, and $p_k=0$ for $k > \xi$. In particular, given $0 < \epsilon < 1$, we define $\xi$ as
\begin{equation*}
    \xi =  \arg \min_{\xi'} \sum_{k=0}^{\xi'} e^{-E_k/T}  > (1-\epsilon) \sum_{l=0}^{\infty} e^{-E_l/T}.
\end{equation*}
Note that $\xi$ depends on $T$ but also on $\{E_k\}$. In general, given a fixed $\epsilon$, $\xi$ increases with the total number of ions $N_A+N_B$, meaning that the amount of noise increases with the system size.

In Fig.~\ref{fig:Paul:traps} we show different settings based on the described trap. In Fig.~\ref{fig:cartoon_1trap}, we assume that module $A$ consists of the first $\lceil K/2 \rceil$ ions of the chain while the rest constitute module $B$. In Fig.~\ref{fig:cartoon_cold_trap}, we consider each module to be in an independent trap. Module $A$ is of an arbitrary size system, while $B$ consists of a single ion. Such setting corresponds to the quantum analog of the cold mediating system described in Sec.~\ref{sec.Cold.mediating.system}, as the trap frequency (which characterizes the strength of the trapping potential) of $A$ is chosen to be one hundred times larger than the frequency of trap $B$. Finally, in Fig.~\ref{fig:cartoon_2traps} we consider both modules to be of the same size and each of them is within an independent trap. In Figs.~\ref{fig:1trap}, \ref{fig:cold_trap} and \ref{fig:2traps}, we plot the optimal infidelity for the mentioned setting for different system sizes. For all cases, we obtain that increasing the number of ions within the traps leads to an enhancement of optimal fidelity for any time $\Delta t$. However, similarly to the case analyzed in Sec. \ref{sec.Independent.noise}, for two independent traps to optimize the logical subspaces does not further enhance the fidelity significantly.

\subsection{General case}
\label{sec.General.case}

In this section, we analyze a more general scenario where $H$ cannot be approximated as diagonal in the basis $\left\{ \ket{E_k} \otimes \ket{\bar{i} \bar{j}} \right\}$. In this case, the position of the physical systems entangles with the logical qubits, which prevents one from interpreting the logical coupling strength as a stochastic variable. Instead, we need to explicitly compute the coherent evolution of the whole system; see Eq.~\eqref{eq:Uquantized}.

\begin{figure}
    \centering
    \subfloat[\centering]{\includegraphics[width=0.8\columnwidth]{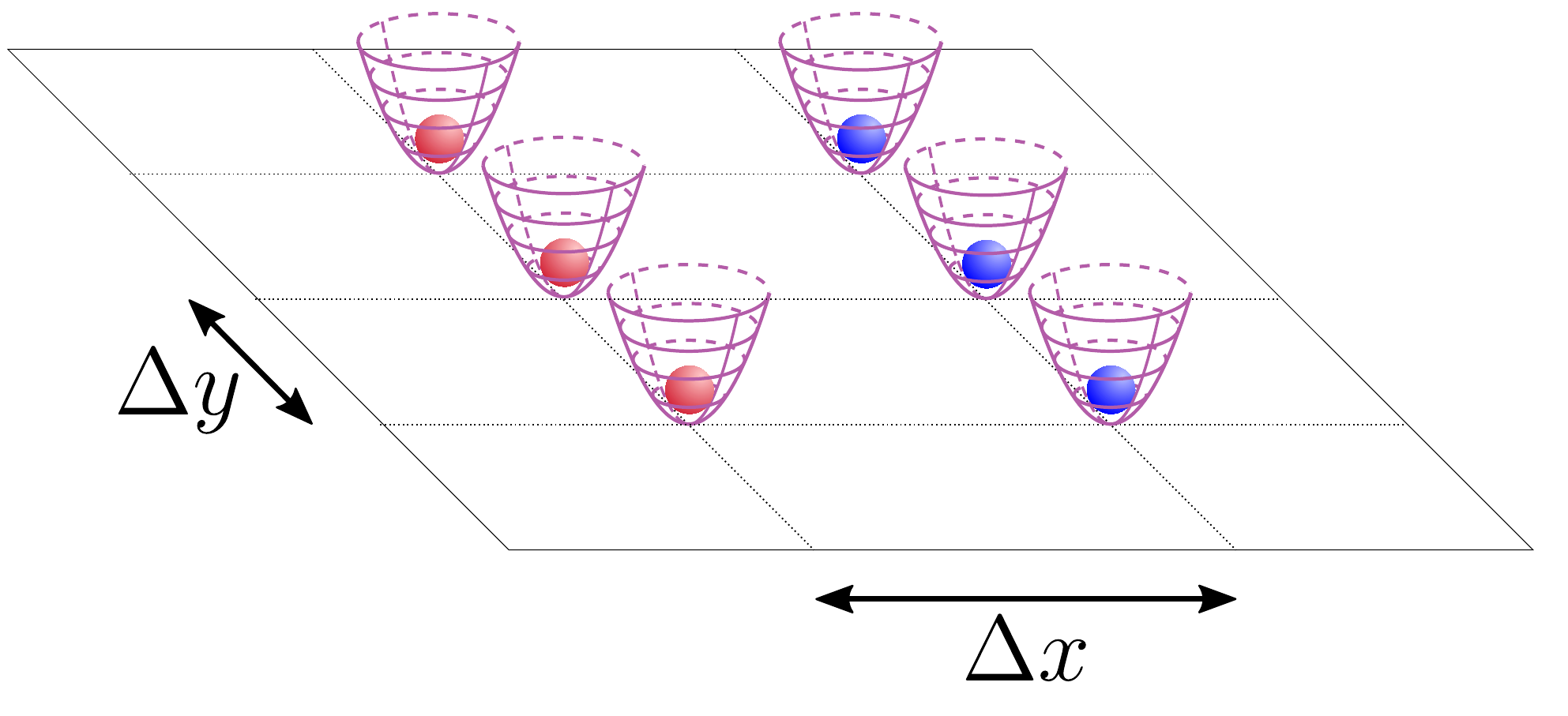}\label{fig:cartoon:2dlattice}} \\
    \subfloat[]{\includegraphics[width=0.735\columnwidth]{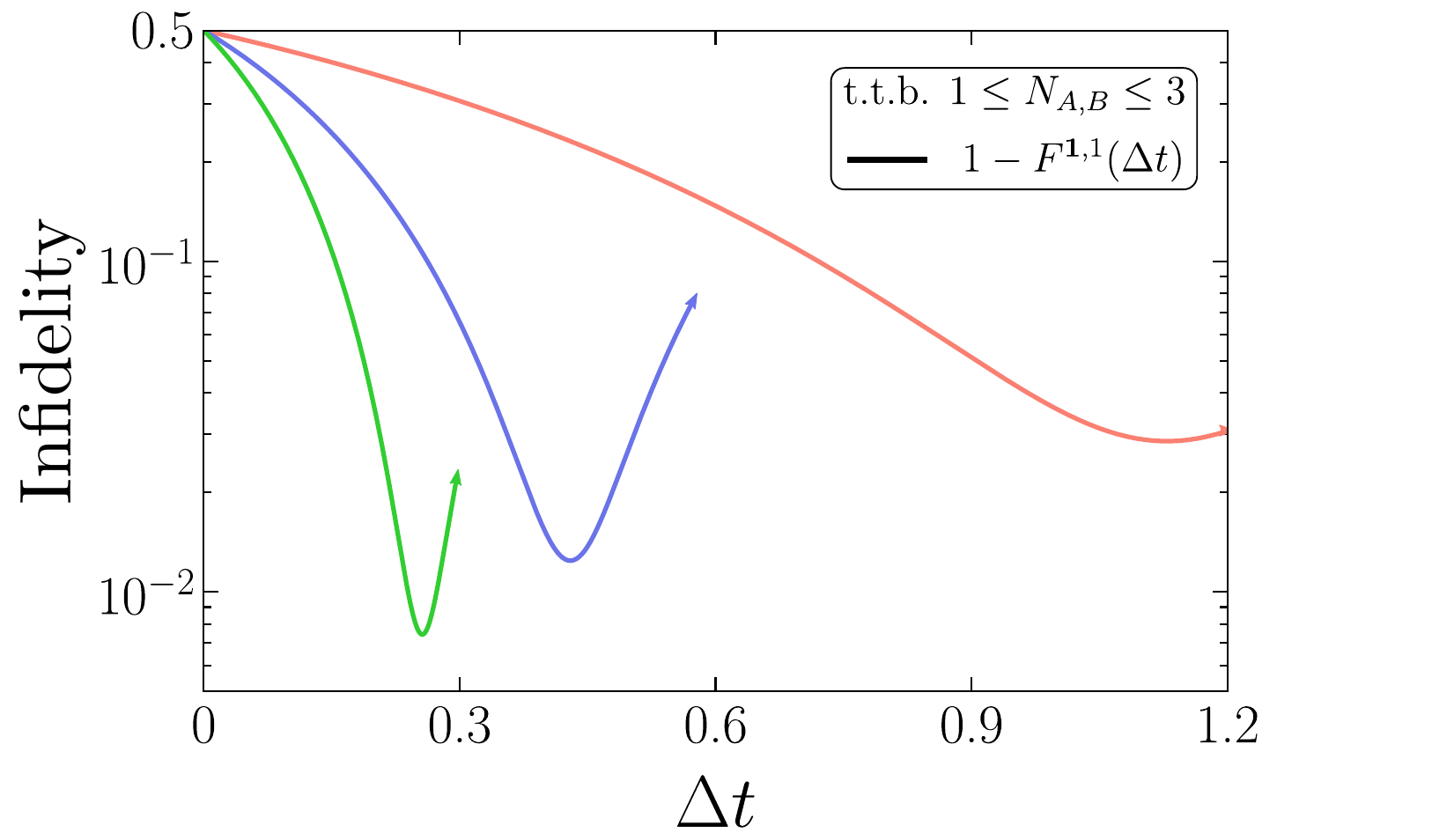}\label{fig:plot:2dlattice}}
    \caption{\label{fig:2dlattice} (a) The physical qubit systems are arranged in a 2D lattice. (b) Infidelity of the implementation of $U = e^{-\ti \frac{\pi}{4} ZZ}$ as a function of time for different system sizes, of the setting shown in (a) with $\omega = 30$, $\Delta x = \Delta y = 2$, $\gamma = 3$, $J = 5$ and $\rho_m = \id/3^{2 N}$.}
\end{figure}

We analyze a particular example of a 2D lattice where each physical system is trapped in an independent harmonic potential, i.e., $H_{\text{m}} = \text{K} + \text{V}$, where K is the kinetic energy of the physical systems and
\begin{equation*}
    \text{V} = \frac{\omega^2}{2} \sum_{i=1}^N \left( \; \left| \vec{r}_i - \vec{r}_i^{\;0} \right|^2 + \left| \vec{q}_i - \vec{q}_i^{\;0} \right|^2 \; \right),
\end{equation*}
is the potential energy where $\vec{r}^{\;0}_i = (0, i \Delta y)$ and $\vec{q}^{\;0}_i = (\Delta x, i \Delta y)$ are the equilibrium positions of the physical systems and $N = N_{A,B}$; see Fig.~\ref{fig:cartoon:2dlattice}. This is a good model for the trapping of neutral atoms, which are not subject to the strong Coulomb interaction.

Due to computational limitations, we cannot consider the infinite spectrum of $H_{\text{m}}$. Instead, we consider the mechanical degree of freedom of each particle to be a three-level system spanned by the ground and the first two excited states, i.e., the Hilbert space of the physical qubit systems positions is given by $\bigotimes_{k=1}^{N} \mathcal{H}_{\text{m}}^{A_k} \otimes \mathcal{H}_{\text{m}}^{B_k}$ where $\mathcal{H}_{\text{m}}^k = span \big\{ \ket{0_k,0_k}, \, \ket{0_k,1_k}, \, \ket{1_k,0_k} \big\}$ and $\ket{ m_k \, n_k}$ is the state of the $k$th particle at the $m(n)$th excited state in the $x(y)$ direction, i.e.,
\begin{equation*}
\begin{gathered}
    \left\langle \, \vec{r}_i \, \left| \, m_{A_i} \, n_{A_i} \right\rangle \right. = \Psi^\omega_m \! \left(r_{xi}-r^0_{xi}\right) \, \Psi^\omega_n\! \left(r_{yi}-r^0_{yi}\right), \\
    \left\langle \, \vec{q}_j \, \left| \, m_{B_j} \, n_{B_j} \right\rangle \right. = \Psi^\omega_m \! \left(q_{xj}-q^0_{xj}\right) \, \Psi^\omega_n \! \left(q_{yj}-q^0_{yj}\right),
\end{gathered}
\end{equation*}
where $\Psi^{\omega}_n(x)$ is the wave function of the $n$th excited state of a quantum harmonic oscillator of frequency $\omega$.

Figure~\ref{fig:plot:2dlattice} shows the infidelity for different system sizes using the trivial encoding, given by $\bo{a} = \bo{b} = \bo{1}$, and if we initialize the mechanical degree of freedom of the system in the maximally mixed state $\rho_{\text{m}} = \id / 9^N$. Due to numeric limitations, we cannot compute the optimal infidelity curve. However, we show that the trivial encoding already significantly enhances the reachable fidelity and implementation.

\section{Background noise}
\label{sec: background noise}

We now show that on top of the error mitigation capabilities against thermal fluctuations, one can also make the system insensitive to a noisy background field. So we consider an extra term in the Hamiltonian of the form
\begin{equation*}
    H_{\text{ext}} = \sum_{k=1}^K h_k \, Z_k
\end{equation*}
where $h_k$ is the intensity of the external field at the position of the $k$th physical qubit system. If $h_k$ is unknown, $H_{\text{ext}}$ yields to a \textit{phase damping channel} acting on each qubit $\bigotimes_{i=k}^K \mathcal{D}_k$ where
\begin{equation*}
\begin{aligned}
    \mathcal{D}_k(\rho) & = \int e^{- \ti \theta Z_k} \, \rho \, e^{\ti \theta Z_k} \, p(\theta) \, \mathrm{d} \theta,
\end{aligned}
\end{equation*}
where if $p(\theta) = \mathcal{N}\!\left[0,\sigma^2\right]$, it is equivalent to a phase flip channel, see Ref. \cite{hot_nielsen2002quantum}.

However, we can easily make the system insensitive to such noise at a certain time $\tau$ by flipping all qubits at $\tau/2$, i.e.,
\begin{equation*}
    X^{\otimes K} e^{- \ti (H_{\text{zz}} + H_{\text{ext}}) \tau/2} \, X^{\otimes K} e^{- \ti (H_{\text{zz}} + H_{\text{ext}}) \tau/2} = e^{- \ti H_{\text{zz}} \tau},
\end{equation*}
where we used that $X_1X_2 \, e^{\ti \theta Z_1Z_2} X_1X_2 = e^{\ti \theta Z_1Z_2}$ and $X_1X_2 \, e^{\ti (h_1 Z_1+ h_2 Z_2)} X_1X_2 = e^{-\ti (h_1 Z_1+ h_2 Z_2)}$. In this way, we can generate the required evolution $e^{-\ti H_{\text{zz}}}$. Then evolving the system under $e^{-\ti H_{\text{zz}}}$ while performing fast flips at a specific time we can effectively generate the evolution with noninteger spin values, as we explain in former work in Ref. \cite{hot_riera2023simulator} Sec. 3.6.

\section{Conclusion and outlook}
\label{sec.conclusions}
                                   
In this article, we introduced a technique to perform high-fidelity logical two-qubit gates where the constituents of the logical system are affected by thermal noise. In particular, we detailed alternative quantum computing architectures based on interconnected modules. We assume full control over the individual modules is given by not between them. To connect the modules one would use the physical interaction between their constituting qubits, which makes the scheme vulnerable to thermal noise. We considered different settings where position noise affects in different manners, including collective and individual noise or classical and fullyquantized qubit position treatments. Due to computational restrictions, we have been forced to simplify the models. However, even with these simplifications our models still resemble realistic setups and we compensate them by the large variety of treated examples. We showed that our method could make the schemes position-noise resilient by encoding logical qubits in the physical ones and using the logical encoding to mitigate the effects of position fluctuations. In particular, we show the fidelity of two-qubit gates can be enhanced by increasing the size of the logical systems. In addition, using logical qubits one can obtain enlarged interaction couplings that reduce the implementation time.

We also show that the setup can be totally protected from other kinds of noise such as background radiation without any extra cost in fidelity or resources.

The central ingredient enabling our approach is fixed distant-dependent Ising-type interactions between physical qubit systems. While such interactions are present to some extent in various systems \cite{hot_Porras2004, hot_richerme_2014non, hot_zhang2017observation, hot_Joshi_2020, hot_Pagano_2020, hot_demille2002quantum, hot_yelin2006schemes, hot_Browaeys_2016, hot_bluvstein2022quantum}, the question of whether they can be used to realize the proposed methods in given setups is left for future work.


\begin{acknowledgments}
This research was funded in whole or in part by the Austrian Science Fund [Fonds zur Förderung der wissenschaftlichen Forschung (FWF)] [Grants DOI: 10.55776/P36009, 10.55776/P36010 and 10.55776/COE1. For open access purposes, the author has applied a CC BY public copyright license to any author accepted manuscript version arising from this submission. Finanziert von der Europ\"aischen Union - NextGenerationEU.
\end{acknowledgments}

\appendix

\section{Fidelity of the \textit{ZZ}-damping channel}
\label{appendix:F}

Here, we derive Eq.~\eqref{eq:F0} for the fidelity of the \textit{ZZ}-damping channel. So on the one hand we consider the unitary entangling gate
\begin{equation*}
    U = e^{-\ti \frac{\pi}{4} ZZ},
\end{equation*}
and on the other hand the \textit{ZZ}-damping channel
\begin{equation}
    \label{eq:quantumchannel}
    \mathcal{U}(\bullet) = \int e^{-\ti \theta ZZ} \bullet \, e^{\ti \theta ZZ} \, p(\theta) \, \mathrm{d}\theta,
\end{equation}
where $p(\theta)$ is the probability function of $\uptheta$.

The Choi fidelity of $\mathcal{U}$ with respect to $U$ is given by $F = \bra{\Phi_U} \Phi_{\mathcal{U}} \ket{\Phi_U}$, where
\begin{align*}
    \ket{\Phi_U} & = \frac{1}{2} \sum_{i,j=0}^1 \ket{ij} \otimes U \ket{ij}, \\
    \Phi_{\mathcal{U}} & = \frac{1}{4} \sum_{i,j,k,l=0}^1 \ket{ij}\!\bra{kl} \, \otimes \mathcal{U}\big( \ket{ij}\!\bra{kl} \big) \\
    & = \frac{1}{4} \int \! \proj{\Phi_\theta} p(\theta) \, \mathrm{d}\theta,
\end{align*}
and
\begin{equation*}
    \ket{\Phi_\theta} = \frac{1}{2} \sum_{i,j=0}^1 \ket{ij} \otimes e^{-\ti \theta ZZ} \ket{ij}.
\end{equation*}
Note that computing
\begin{align*}
    \braket{\Phi_U}{\Phi_\theta} & = \frac{1}{4} \sum_{i,j=0}^1 \bra{ij} e^{\ti (\frac{\pi}{4} - \theta) ZZ} \ket{ij} \\
    & = \frac{1}{4} \sum_{i,j=0}^1 e^{\ti (-1)^{i+j} (\frac{\pi}{4} - \theta)} \\
    & = \frac{1}{2} \left[ e^{\ti (\frac{\pi}{4} - \theta)} + e^{-\ti (\frac{\pi}{4} - \theta)} \right] \\
    & = \cos \! \left( \tfrac{\pi}{4} - \theta \right) ,
\end{align*}
the fidelity can be written as
\begin{align*}
    F & = \int \left| \braket{\Phi_U}{\Phi_\theta} \right|^2 p(\theta) \, \mathrm{d}\theta \\ 
    & = \int \cos^2 \! \left( \tfrac{\pi}{4} - \theta \right) p(\theta) \, \mathrm{d}\theta \\
    & = \left\langle \cos^2 \! \left( \tfrac{\pi}{4} - \uptheta \right) \right\rangle.
\end{align*}
Note that in the main text, we consider $\theta = \mu \, \Delta t$.

\section{Cold mediating system final fidelity}
\label{Appendix:F1F2}

We consider the following sequence of gates such that it mediates an interaction between the first and the second qubit,
\begin{equation}
\begin{aligned}
    (Z_1 Z_2)^k \, P^{(k)}_0 \, e^{\ti \alpha X_0} \, \text{CZ}_{01} \text{CZ}_{02} \ket{+}_0 \ket{\psi}_{12} & \\
    \mapsto \, e^{\ti \alpha ZZ} \ket{\psi}_{12} & ,
\end{aligned}
\end{equation}
where $P^{(k)} = \proj{k}$, and $\ket{\psi}$ is an arbitrary state.

Then if we use the noisy interaction to implement the CZ gates, the channel describing the process is given by
\begin{equation*}
    \mathcal{S}(\bullet) = \int \mathcal{S}^{\alpha \beta}(\bullet) \, p(\alpha) \, p(\beta) \, \mathrm{d}\alpha \, \mathrm{d}\beta  
\end{equation*}
where
\begin{equation*}
\begin{aligned}
    \mathcal{S}^{\alpha \beta}(\bullet) = \sum_{k=0}^1 M^{(k)}_{012} \, U_{01}^{( \alpha)} \, U_{01}^{( \beta)} (\bullet) \,U_{01}^{( \alpha) \, \dagger} \, U_{01}^{( \beta)\, \dagger} \, M^{(k) \, \dagger}_{012},
\end{aligned}
\end{equation*}
with $U^{(\alpha)} = e^{-\ti \alpha ZZ}$ and $M^{(k)} = (Z_1 Z_2)^k \bra{k}_0 e^{-\ti \frac{\pi}{4} X_0} \, e^{\ti \frac{\pi}{4} (2 Z_0+Z_1+Z_2)}$.

The fidelity for fixed but arbitrary angles $\alpha$ and $\beta$
\begin{equation*}
     F_{\alpha\beta} = \bra{\Phi_U} \Phi_{\alpha \beta} \ket{\Phi_U} = \cos^2\!\left( \tfrac{\pi}{4} - \alpha\right) \cos^2\!\left(\tfrac{\pi}{4} - \beta \right),
\end{equation*}
where $\Phi_{\alpha \beta}$ is the Choi state of $\mathcal{S}^{\alpha \beta}$. Therefore the fidelity of $\mathcal{S}$ is given by the expected value of $F_{\alpha \beta}$, i.e.,
\begin{align*}
    F & = \int F_{\alpha \beta} \, p(\alpha) \,p(\beta) \, \mathrm{d}\alpha \, \mathrm{d}\beta \\ 
    & = \int \! \cos^2 \! \left( \tfrac{\pi}{4} - \alpha \right) \cos^2 \! \left( \tfrac{\pi}{4} - \beta \right) p(\alpha) \, p(\beta) \, \mathrm{d}\alpha \, \mathrm{d}\beta \\
    & = \left\langle \cos^2 \! \left( \tfrac{\pi}{4} - \upalpha \right) \right\rangle \left\langle \cos^2 \! \left( \tfrac{\pi}{4} - \upbeta \right) \right\rangle = F_2 \, F_3,
\end{align*}
where $F_j$ is the fidelity of the noisy control gate between the $j$th hot qubit and the auxiliary system.

\section{Cold mediating system: 2D}
\label{sec.Cold.mediating.system.2D}

Here we describe a direct 2D extension of the example analyzed in Sec.~\ref{sec.Cold.mediating.system}. We assume the position of the qubits in $A$ are given by
\begin{equation}
\begin{aligned}
    \vec{r}_1 & = (\Delta x, \Delta y, 0),  \qquad & \vec{r}_6 & = (2\Delta x, 2\Delta y, 0),    \\
    \vec{r}_2 & = (2\Delta x, \Delta y, 0), \qquad & \vec{r}_7 & = (0, 0, 0), \\
    \vec{r}_3 & = (\Delta x, 2\Delta y, 0), \qquad & \vec{r}_8 & = (0, 2\Delta y, 0),   \\
    \vec{r}_4 & = (0, \Delta y, 0),         \qquad & \vec{r}_9 & = (2\Delta x, 0, 0),   \\
    \vec{r}_5 & = (\Delta x, 0, 0),
\end{aligned}
\end{equation}
and the position of the qubit in $B$ is a stochastic variable given by $\vec{\text{q}} = (\text{q}_{1x}, \text{q}_{1y}, q_{1z})$ where $\text{q}_{1x} \sim \mathcal{N}\!\left[ \Delta x, \sigma^2 \right]$, $\text{q}_{1y} \sim \mathcal{N}\!\left[ \Delta y, \sigma^2 \right]$ and $q_{1z} = \Delta z$.

In Fig.~\ref{fig:2D:froz}, we plot the infidelity as a function of time for different system sizes. We obtain a behavior similar to the 1D case. However, in this case, the physical qubit systems are not located in a symmetric way leading to asymmetric fidelity improvement with the system size.

\section{Collective position noise: 2D}
\label{sec.collective.position.noise.2D}

\begin{figure}
    \centering
    \subfloat[]{\includegraphics[width=0.66\columnwidth]{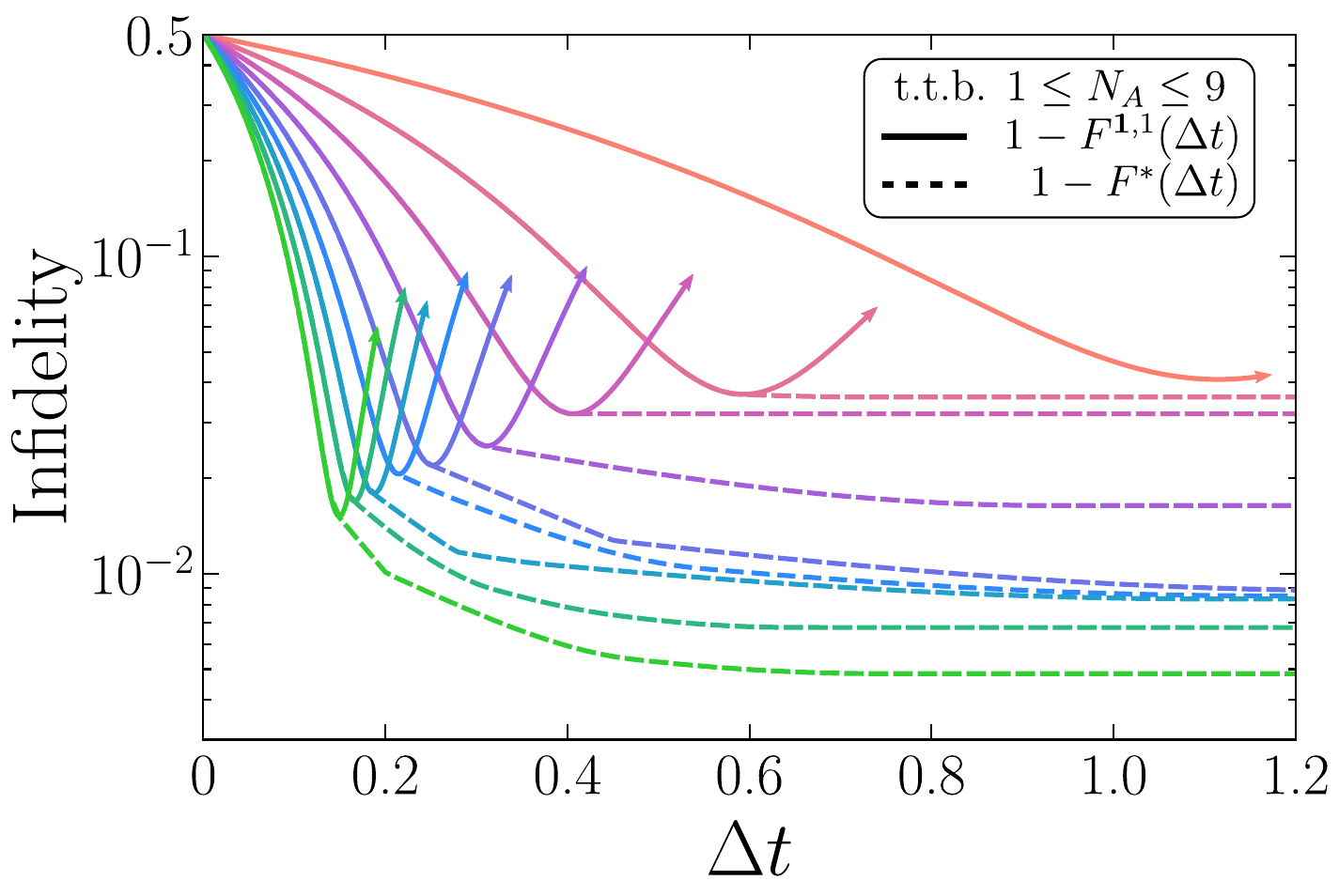}\label{fig:2D:froz}} \\
    \subfloat[]{\includegraphics[width=0.66\columnwidth]{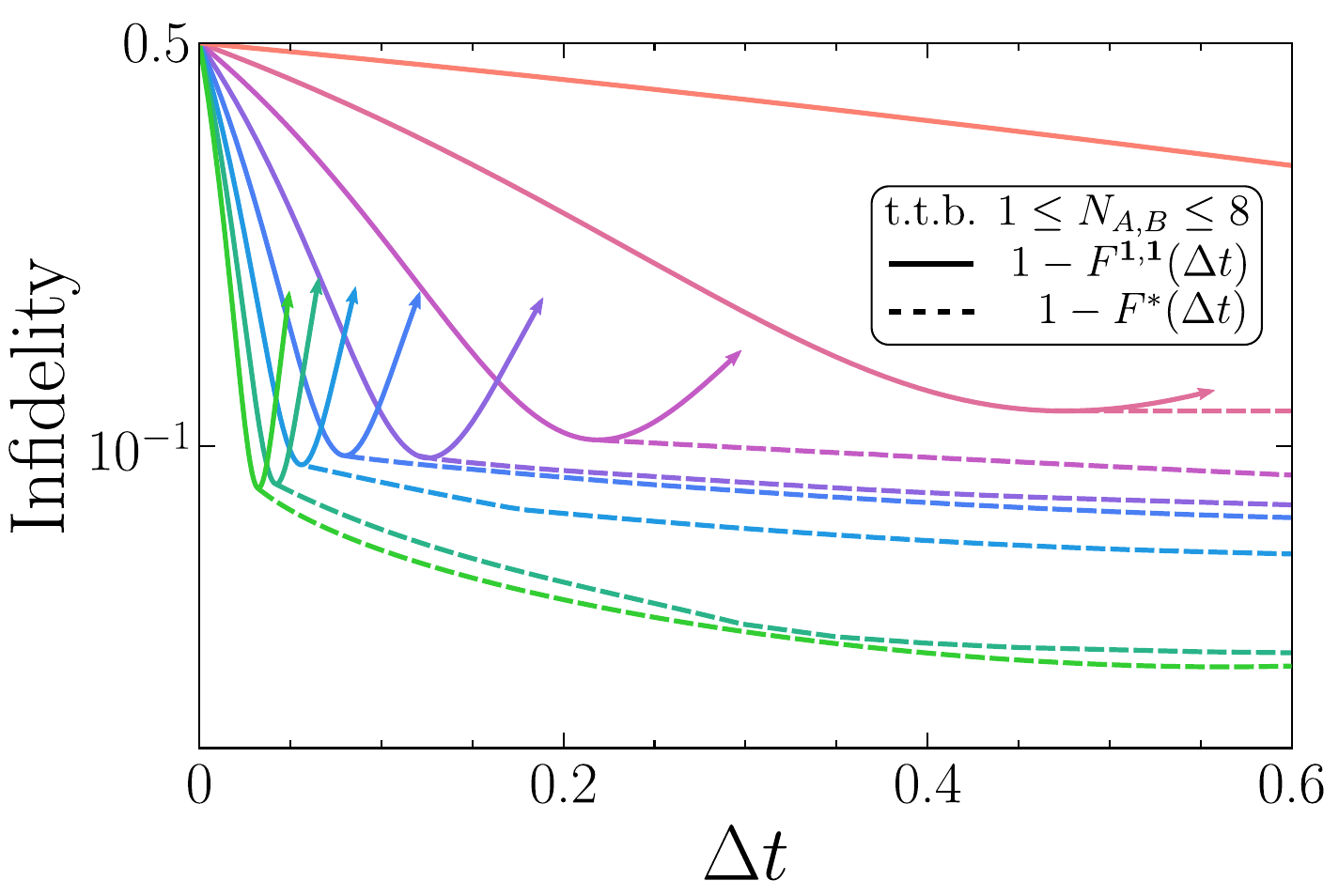}\label{fig:2D:col}}
    \caption{\label{fig:2D:appendix} (a) Optimal infidelity as a function of time for different values of $N_A$; with $N_B = 1$, $\gamma=1$, $\Delta x = \Delta y = \Delta z = 1$ and $\sigma = 1$. (b) Optimal infidelity as a function of time for different values of $N_A$ and $N_B$, with $\gamma = 1$, $\Delta x = \Delta z = 1$ and $\sigma = 2$.}
\end{figure}

Here we describe a direct 2D extension of the example analyzed in Sec.~\ref{sec.Collective.position.noise}. We assume the position of the physical qubit system is given by
\begin{align*}
    \vec{\textbf{r}} & = \vec{\bo{r}}^{\,0} + (\vec{\text{r}}, \vec{\text{r}}, \dots, \vec{\text{r}}\,), \\
    \vec{\textbf{q}} & = \vec{\bo{q}}^{\,0} + (\vec{\text{q}}, \vec{\text{q}}, \dots, \vec{\text{q}}\,),
\end{align*}
where $\vec{\text{r}}$ and $\vec{\text{q}} \sim \mathcal{N}\!\left[0,\sigma^2\right]$, and where in this particular example $\vec{\bo{r}\;}^0$ is given by
\begin{equation}
\begin{aligned}
    \vec{r}^{\;0}_1 & = (0, 0, \Delta z),         \qquad & \vec{r}^{\;0}_5 & = (\Delta x, \Delta y, \Delta z),   \\
    \vec{r}^{\;0}_2 & = (\Delta x, 0, \Delta z),  \qquad & \vec{r}^{\;0}_6 & = (2 \Delta x, \Delta y, \Delta z), \\
    \vec{r}^{\;0}_3 & = (2\Delta x, 0, \Delta z), \qquad & \vec{r}^{\;0}_7 & = (0, 2 \Delta y, \Delta z),        \\
    \vec{r}^{\;0}_4 & = (0, \Delta y, \Delta z),  \qquad & \vec{r}^{\;0}_8 & = (\Delta x, 2 \Delta y, \Delta z)  \\
\end{aligned}
\end{equation}
and $\vec{\bo{q}\;}^0$ by
\begin{equation}
\begin{aligned}
    & \vec{q}^{\;0}_1 = (0, 0, 0),           \qquad & \vec{q}^{\;0}_5 & = (\Delta x , \Delta y , 0),  \\
    & \vec{q}^{\;0}_2 = (\Delta x , 0, 0),   \qquad & \vec{q}^{\;0}_6 & = (2\Delta x , \Delta y , 0), \\
    & \vec{q}^{\;0}_3 = (2 \Delta x , 0, 0), \qquad & \vec{q}^{\;0}_7 & = (0, 2 \Delta y , 0),        \\
    & \vec{q}^{\;0}_4 = (0, \Delta y , 0),   \qquad & \vec{q}^{\;0}_8 & = (\Delta x , 2 \Delta y , 0).
\end{aligned}
\end{equation}
In Fig.~\ref{fig:2D:col} we plot the infidelity as a function of time for different system sizes. Like in the previous section, here we obtain a similar behavior to the 1D case. However, in this case, the physical qubit systems are also not located in a symmetric way leading to asymmetric fidelity improvement with the system size.

\section{Trap modes}
\label{appendix:trapmodes}

We consider a chain of $K$ ions in a 1D trap as described in Ref. \cite{hot_james1997quantum}. The ions are assumed to be strongly bound in the $y$ and $z$ directions but weakly bound in a harmonic potential in the $x$ direction. The motion of each ion will be influenced by an overall harmonic potential due to the trap electrodes and by the Coulomb force exerted by all of the other ions. Hence, the physical qubit systems are subjected to the potential
\begin{equation*}
    \text{V} = \frac{\omega^2}{2} \sum_{m = 1}^K x_m^2 + \sum_{1\leq i<j \leq K} \frac{\chi}{|x_i - x_j|}
\end{equation*}
where $\omega$ is the trap frequency, and $\chi>0$ is a constant related to the Coulomb interaction and the ionization of the ions.

The equilibrium positions, $\bo{x}^0 = (x_1^0,\dots, x^0_K)$, are such that fulfill
\begin{equation}
    \label{eq:V0}
    \left( \frac{\partial \text{V}}{\partial x_i} \right)_{\bo{x}^0} = 0,
\end{equation}
for $1 \leq i \leq K$. Note if one writes Eq.~\eqref{eq:V0} in terms of the dimensionless position coordinates given by $\bar{x}_i =  x_i/L$ where $L^3 =  \chi/\omega^2$, then in Eq.~\eqref{eq:V0} the only parameter left is $K$, e.g., for $K=3$ one obtains
\begin{equation*}
    \bar{x}^0_1 = - \left( \frac{5}{4} \right)^{1/3}, \quad
    \bar{x}^0_2 = 0, \quad \bar{x}^0_3 = \left( \frac{5}{4} \right)^{1/3}.
\end{equation*}
Therefore, given any $\omega$ and $\chi$ we can find the equilibrium positions as $x^0_i = L \, \bar{x}^0_i$.

If the ions are cold enough to assume small displacements around the equilibrium positions, one can perform a Taylor expansion of the potential around $\bo{x}^0$ up to the second order and obtain the following approximation of the Hamiltonian
\begin{equation*}
    \label{eq:Hmec}
    H_{\text{m}} \approx \frac{1}{2} \Big( \, |\dot{\bo{x}}|^2 + \omega^2 \, \delta \bo{x} \cdot \textbf{V}'' \cdot \delta \bo{x} \, \Big),
\end{equation*}
where $\delta \bo{x} =  \bo{x} - \bo{x}^0_i$ collects the displacement from the equilibrium position of the ions and $\textbf{V}'' = \frac{1}{\omega^2} \left[ \partial^2 \text{V}/(\partial x_i \partial x_j)\right]_{\bo{x}^0}$ is a parameter-independent matrix, i.e., independent of $\omega$ and $\chi$. Note $\bo{V}''$ is symmetric, and hence its eigenvectors form an orthonormal basis; i.e., there exists a set of vectors $\left\{\bo{v}^{(m)}\right\}$ such that $\textbf{V}''\cdot \bo{v}^{(m)} = \lambda_m \bo{v}^{(m)}$ and $\bo{v}^{(m)} \cdot \bo{v}^{(n)} = \delta_{mn}$. Therefore, one can write $\delta \bo{x}$ as a linear combination of the eigenvectors, i.e.,
\begin{equation*}
    \delta \bo{x} = \sum_{m=1}^K u_m \bo{v}^{(m)},
\end{equation*}
where $u_m = \bo{v}^{(m)} \cdot \delta \bo{x}$, and write the Hamiltonian in the following coordinates
\begin{equation*}
    H_{\text{m}} \approx \sum_{m=1}^K \frac{1}{2} \left( \dot{u}_m^2 + \omega^2 \, \lambda_m \, u_m^2 \right).
\end{equation*}
Therefore, in the new coordinates $\{u_m\}_{m=1}^K$, e.g., for $K = 3$
\begin{align*}
    u_1(\delta\bo{x}) & = \frac{1}{\sqrt{3}} (\delta x_1 + \delta x_2 + \delta x_3), \\
    u_2(\delta\bo{x}) & = \frac{1}{\sqrt{2}} (\delta x_1 - \delta x_3),              \\
    u_3(\delta\bo{x}) & = \frac{1}{\sqrt{6}} (\delta x_1 - 2\delta x_2 + \delta x_3),
\end{align*}
the Hamiltonian decouples in $K$ independent harmonic oscillators of frequency $\nu_m = \sqrt{\lambda_m} \, \omega$ where $\lambda_1 = 1$, $\lambda_2 = 3$ and $\lambda_3 = 29/ 5$.

Therefore, the eigensystem of $H_{\text{m}}$ is given by $\{ E_{\bo{m}}, \ket{\bo{m}} \}$, where $\ket{\bo{m}} = \bigotimes_{i=1}^K \ket{m_i}$ contains $m_k$ excitations or phonons on the $k$th mode. In other words,
\begin{equation*}
    \ket{m_i} = \int \Psi^{\nu_i}_{m_i} (u_i) \ket{u_i} \mathrm{d}u_i,
\end{equation*}
where
\begin{equation*}
    \Psi^{\nu}_n (x) = \frac{1}{\sqrt{2^n n!}} \left( \frac{\nu}{\pi}\right)^{1/4} e^{- \nu x^2/2} \, \text{H}_n \! \left( \sqrt{\nu} \, x \right)
\end{equation*}
is the wave function of the $m$th excited state of a quantum harmonic oscillator of frequency $\nu$ and $\text{H}_m(x)$ are the Hermite polynomials, and hence 
\begin{equation*}
    E_{\bo{m}} = \sum_{i=1}^K \nu_i \left( m_i + \frac{1}{2} \right).
\end{equation*}

In Sec.~\ref{sec.traps} of the main text we consider different settings based on 1D Paul traps. Here, we provide more detail on the considered setting:
\begin{itemize}
    \item One single Paul trap. We consider modules $A$ and $B$ within the same 1D Paul trap; see Fig.~\ref{fig:cartoon_1trap}. Module $A$ consist of the first $N_A = \lceil K/2 \rceil$ physical qubit systems, i.e., $\vec{r}_i =(x_i,0)$, while the rest $N_B = \lfloor K/2 \rfloor$ constitute module $B$, i.e., $\vec{q}_j =(x_{N_A+j},0)$. Therefore, the wave function of the $\ket{\bo{m}}$ is given by
    \begin{equation*}
        \Phi_{\bo{m}}(\bo{r},\bo{q}) = \prod_{i=1}^{N_A+N_B} \Psi^{\nu_i}_{m_i} \! \left[u_i\left(\delta\bo{r},\delta\bo{q}\right)\right]
    \end{equation*}
    
    \item Two independent 1D Paul traps. We then consider each module in an independent trap; see Figs. \ref{fig:1trap} and \ref{fig:2traps}. The two traps are oriented in the same direction and separated by a distance $\Delta y$. Each trap has a certain frequency $\omega_A$ and $\omega_B$. In addition, we consider the ions in each trap to have a different ionization $\chi_A$ and $\chi_B$ such that both traps have the same equilibrium positions, i.e., $r_{xi}^0 = q_{xi}^0$. This is achieved by tuning the ionization of the ions to such that it is fulfilled that $L_A = L_B$.

    Therefore, in this case, the mechanical Hamiltonian is the sum of the Hamiltonian of each trap, i.e., $H_{\text{m}} = H_{\text{m}}^A + H_{\text{m}}^B$, and the eigenstates are given by $\ket{\bo{m}}_A\ket{\bo{n}}_B$. The wave functions of the eigenstates depend on the equilibrium positions and the natural frequencies of each trap, $\omega_A$ and $\omega_B$. If $\nu_m = \sqrt{\lambda_m} \, \omega_A$ and $\tilde{\nu}_m = \sqrt{\lambda_m} \, \omega_B$ then
    \begin{equation*}
        \Phi_{\bo{m n}}(\bo{r},\bo{q}) = \prod_{i=1}^{N_A} \Psi^{\nu_i}_{m_i} [u_i(\delta\bo{r})] \;\; \prod_{j=1}^{N_B} \Psi^{\tilde{\nu}_j}_{n_j} [u_j(\delta\bo{q})].
    \end{equation*}
\end{itemize}

\begin{figure}[h!]
    \centering
    \subfloat[]{\includegraphics[width=\columnwidth]{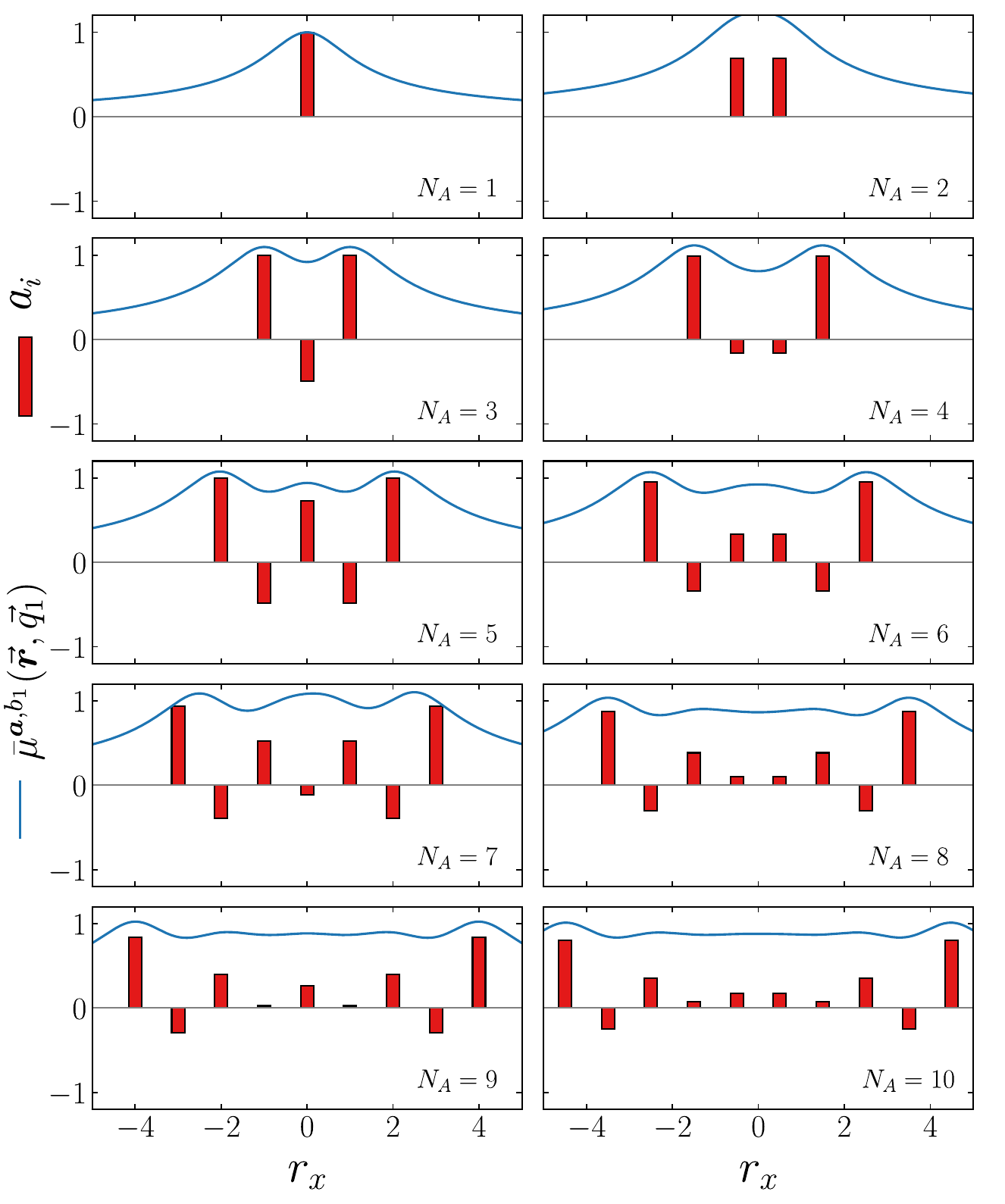} \label{appendix.fig:A}} \\
    \subfloat[]{\includegraphics[width=\columnwidth]{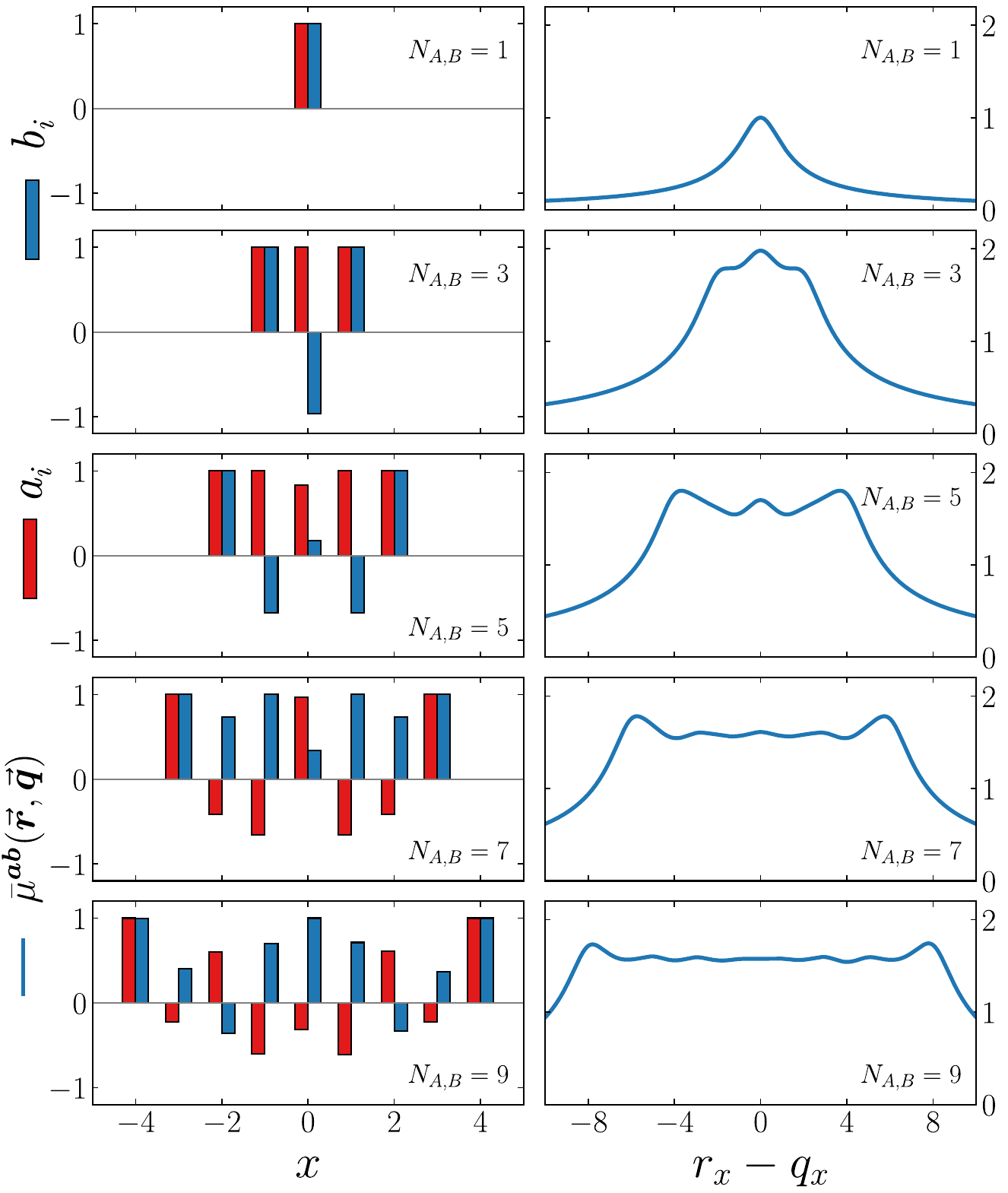} \label{appendix.fig:AB}}
    \caption{(a) Optimal logical subspace $\bo{a}$ for $t = 0.9$, for the scheme details in Fig.~\ref{Fig:frozen_1D} in the main text. The $i$th bar of the histogram corresponds to the value of $a_i$. (b) Optimal logical subspace $\bo{a}$ for $t = 0.5$, for the scheme details in Fig.~\ref{Fig:collective_1D} in the main text. The $i$th bar of the histogram corresponds to the value of $a_i$.}
\end{figure}

\section{Optimal logical subspaces}
\label{appendix:optimal.logical.subspaces}

We find in this section the optimal values of $\bo{a}$ and $\bo{b}$ for the scenarios analyzed in Secs. \ref{sec.Cold.mediating.system} and \ref{sec.Collective.position.noise}; see Figs.~\ref{appendix.fig:A} and \ref{appendix.fig:AB}, respectively. A systematic observation is that the effective spin values are symmetric along the chain due to the setting symmetry. A second observation is that the ``extremal'' qubits (i.e., those farther away from the center) have the larger spin value. The intuition behind this is that this leads to a ``flatter'' potential, which is less sensitive to position noise.

\bibliography{Hot_qubits.bib}
\end{document}